\documentclass[twoside,draft,reqno,12pt]{amsart}
\usepackage[english]{babel}
\usepackage{amsmath}
\usepackage{amssymb,ulem}
\usepackage{enumerate}

\usepackage{amsfonts}
\usepackage{graphicx}

\usepackage[ citecolor=black, linkcolor=black,
colorlinks, hypertexnames=false, plainpages=false]{hyperref}

\usepackage{color}

\newtheorem{lemma}     {Lemma}[section]
\newtheorem{thm}   [lemma]{Theorem}
\newtheorem{teorema1}   [lemma]{Theorem}
\newtheorem{prop}       [lemma]{Proposition}
\newtheorem{coro}       [lemma]{Corollary}
\newtheorem{cong1}      [lemma]{Conjecture}
\newtheorem{remark1}    [lemma]{Remark}
\newtheorem{defin}      [lemma]{Definition}

\numberwithin{equation}{section}

\newcommand{\dis}{\displaystyle}

\newcommand{\arctanh}{\mathrm{arctanh}\,}

\begin{document}

\title[Action minimization and interface motion under forced displacement]
{Action minimization and macroscopic interface motion under forced displacement}

\author{P. Birmpa}
\address{Panagiota Birmpa,
Department of Mathematics, University of Sussex \newline
\indent Brighton, U.K.}
\email{P.Birmpa@sussex.ac.uk}

\author{D. Tsagkarogiannis}
\address{Dimitrios Tsagkarogiannis,
Department of Mathematics, University of Sussex \newline
\indent Brighton, U.K.}
\email{D.Tsagkarogiannis@sussex.ac.uk}

\begin{abstract}
We study an one dimensional model where an interface is the stationary
solution of a mesoscopic non local evolution equation which has been derived by a
microscopic
stochastic spin system. Deviations from this evolution equation can be
quantified by obtaining the large deviations cost functional from the underlying stochastic
process.
For such a functional, derived in a companion paper, 
we investigate the optimal way for a macroscopic interface 
to move from an initial to a final position distant by $R$ within fixed time $T$.
We find that for small values of $R/T$
the interface moves with a constant speed, while for larger values there
appear nucleations of the other phase ahead of the front.
\end{abstract}


\maketitle

\section{Introduction}

In recent years, there has been a significant effort to derive deterministic models
describing two-phase materials and their dynamical properties, \cite{halperin}.
Furthermore, with the inclusion of stochastic effects \cite{FJ} one can study richer
phenomena such as dynamic transitions between local minima. 
This is an extension of ideas already developed in 
the Freidlin-Wentsell theory \cite{FW} on random perturbation of dynamical systems.
Such effects, can
be encoded to action functionals whose minimizers prescribe the optimal transition.
The choice of the action functional is not straightforward. The purpose of this paper
and of the companion \cite{BDT}, is to show that 
given the {\it mesoscopic} deterministic partial differential equation (PDE), one can consider the underlying 
{\it microscopic} stochastic process
(whose scaling limit is the given PDE) and calculate
the corresponding large deviations functional which would provide the
action functional we are after.
This is a well developed idea also in the more general setting of nonequilibrium systems
\cite{BDGJL}
and here we examine it in the context of reversible dynamics describing {\it macroscopic} interface motion.
Furthermore, this connection to the underlying stochastic process
is also insightful for calculating the minimizers.
For example, in the present work we borrow concepts from statistical mechanics
such as contours, free energy, local equilibrium 
which allow us to better understand the structure of the cost functional and hence reduce
it in a simpler and more easily treatable form. 

Similar results have been obtained in the context of the stochastic Allen-Cahn equation.
In \cite{KORV, KRT} the authors study the same problem for $d=1$ while in \cite{MR, BBP2} it is extended to
$d=2,3$. In particular, in \cite{BBP2}
the limit considered is a joint sharp interface and small noise, but the starting point
is at the mesoscopic scale, even though noise is also involved. Some numerical results were also
presented in \cite{ERV}.
In this context, our contribution in this and the companion paper \cite{BDT} is that we 
derive (and subsequently minimize) the large deviations action functional directly from a microscopic process, hence 
completing this program of connecting the three scales: microscopic (process), mesoscopic (equation) and macroscopic (sharp-interface). 
However, for technical reasons we have to restrict ourselves in $d=1$ 
even though several partial results are valid also in higher dimensions.
Note also that in a coarse-grained (almost mesoscopic) scale, we have an equation which is comparable to
a non-local Allen-Cahn type equation
with a noise which is a martingale generated by the microscopic noise of each spin.
On the other hand, in the 
stochastic Allen-Cahn one adds by hand a ``mesoscopic"
white-noise in one dimension, or a properly coloured noise in higher dimensions
(for more details about the motivation see the introduction in \cite{BBP1}).
The connection to the stochastic Allen-Cahn 
is particularly interesting also in view of the results \cite{BPRS, MW} connecting 
the fluctuations of this microscopic process
to the stochastic Allen-Cahn equation in a critical regime. 
We conclude mentioning that the meso-to-macro limit for a closely related evolution equation has been already addressed in \cite{ddp}, but for a postulated action functional given by the $L^{2}$ norm of an external
force corresponding to the deviating profiles.
In fact, we show that the large deviations functional gives
a softer penalization on deviating profiles than
the $L^2$ norm considered in \cite{ddp}, hence our task here is a bit harder and
we need to properly adjust the proof of \cite{ddp} in the
new context.

\section{The model and the main result}

We work in the context of a nonlocal evolution equation which can be
derived by an interacting particles system of Ising spins with Kac interaction and Glauber dynamics,
\cite{CE} and \cite{DOPT}:
\begin{eqnarray}\label{eqn}
\frac{d}{dt}m  =  -m+\tanh\{\beta(J\ast m)\}, 
\qquad
m(0,x)  =  m_0(x),
\end{eqnarray}

where $J\ast m(x,t)=\int_{\mathbb R} J(x-y)m(y,t)\, dy$
and $J\in C^2(\mathbb{R})$ is even,
$J(r)=0$ for all $|r|>1$, $\int_{\mathbb R}J(r)dr=1$ and non increasing for $r>0$.
We also suppose $\beta>1$.
Furthermore, this equation is related to the gradient flow of the free energy functional
      \begin{equation}
    \label{n2.6}
 \mathcal F(m)= \int_{\mathbb R} \phi_\beta(m) dx + \frac 14
 \int_{\mathbb R\times \mathbb R} J(x,y)[m(x)-m(y)]^2dx\,dy,
     \end{equation}
where $\phi_\beta(m)$ is the ``mean field excess free energy''
      \begin{equation*}
\phi_\beta(m) = \tilde\phi_{\beta}(m) - \min_{|s|\le 1}
\tilde\phi_{\beta}(s),\qquad \tilde \phi_{\beta}(m) =
-\frac{m^2}{2} - \frac 1 \beta {\mathcal S}(m),\qquad \beta>1,
     \end{equation*}
     and ${\mathcal S}(m)$ the entropy:
      \begin{equation*}
 {\mathcal S}(m)= - \frac{1-m}{2} \log\,\frac{1-m}{2} -
\frac{1+m}{2}\log\,\frac{1+m}{2}.
     \end{equation*}
We also denote by
\begin{equation}\label{gradient}
f(m):=\frac{\delta\mathcal{F}}{\delta m}=-J\ast m+\frac{1}{\beta}\arctanh m
\end{equation}
the functional derivative of $\mathcal F$.
Thus, the functional in \eqref{n2.6} is a Lyapunov functional for the equation \eqref{eqn}:
\[
\frac{d}{dt}\mathcal F(m)=-\frac{1}{\beta}\int_{\mathbb R} (- \beta J\ast m+\arctanh m)(m-\tanh (\beta J\ast m))\, dx\leq 0,
\]
since the  two factors inside the integral have the same sign.
This structure will be essential in the sequel, e.g. in Theorem~\ref{nucl}.

Concerning the stationary solutions of the equation \eqref{eqn} in $\mathbb R$, it has been proved that 
the two constant functions  $m^{(\pm)}(x):= \pm m_\beta$, with
$m_\beta>0$  solving the mean field equation $\dis{m_\beta=
\tanh\{\beta m_\beta\}}$ are stationary
solutions of \eqref{eqn} and are interpreted as the two pure
phases of the system with positive and negative magnetization.

Interfaces, which are the objects of this paper, are made up from particular 
stationary solutions of \eqref{eqn}. Such solutions, called {\it instantons}, exist for any
$\beta>1$ and we denote them by $\bar
m_\xi(x)$, where $\xi$ is a parameter called the center of the instanton. 
Denoting $\bar m:=\bar m_0$, we have that
     \begin{equation}
          \label{transl}
\bar m_\xi(x)=\bar m(x-\xi),
     \end{equation}
where the instanton $\bar m$ satisfies
     \begin{equation}
     \label{instanton}
\bar m (x)= \tanh\left\{\beta J *\bar m (x)\right\}, \quad x\in
\mathbb R.
     \end{equation}
It is an increasing, antisymmetric function which converges
exponentially fast to $\pm m_\beta$ as $x\to \pm \infty$, see e.g. 
\cite{DOPT5}, and 
there are $\alpha$ and $a$ positive so that
     \begin{equation}
     \label{1.222}
\lim_{x\to \infty} e^{\alpha x}\bar m'(x)=a,
     \end{equation} see \cite{DOP}, Theorem 3.1.
Moreover,  any other solution of \eqref{instanton} which is
strictly positive [respectively negative] as $x\to \infty$
[respectively $x\to - \infty$], is  a translate of $\bar
m(x)$, see \cite{DOPT6}.
Note also that in the case of finite volume $[-\epsilon^{-1}L,\epsilon^{-1}L]$ 
the solution $\bar m^{(\epsilon)}$
with Neumann boundary conditions is 
close to $\bar m$: for every $\epsilon>0$ we consider the non-local mean field equation 
\begin{equation}\label{finitenonlocal}
m^{(\epsilon)}=\tanh\{\beta J^{\mathrm{neum}}\star m^{(\epsilon)}\},\;\;\; |x|\leq\epsilon^{-1}L,
\end{equation}
where $m^{(\epsilon)}\in L^{\infty}([-\epsilon^{-1}L,\epsilon^{-1}L];[-1,1])$ and \[J^{\mathrm{neum}}(x,y):=J(x,y)+J(x,R_{\epsilon^{-1}L}(y))+J(x,R_{-\epsilon^{-1}L}(y)),
\] 
with $R_{l}(y):=l-(y-l)$ being the reflection of $y$ around $l$. By following \cite{BDP3}, Section 3, or \cite{BDP}, Section 3.3, given $\zeta>0$ there exists $\epsilon_0$ such that for every $\epsilon<\epsilon_0$, there is $\bar m^{(\epsilon)}$ which is antisymmetric, solves \eqref{finitenonlocal}, satisfies
\begin{equation}\label{finitevol}
\|\bar m^{(\epsilon)}-\bar m\|_{L^{\infty}([-\epsilon^{-1}L,\epsilon^{-1}L])}<\zeta
\end{equation}
and it is unique in the above neighbourhood. See also \cite{errico}, section 6.2.3.

Hence, if we start with an instanton, the evolution \eqref{eqn} will not move it.
So, in order to impose a speed to the interface one has to add an external force to
the equation \eqref{eqn}. 
The result would be a deviation from \eqref{eqn} and any such deviation $\{\phi(x,t)\}_{x,t}$
corresponds to an external force that can produce it and
which is given by
\begin{equation}\label{functionb}
b(\phi)(x,t):=\dot{\phi}(x,t)+\phi(x,t)-\tanh(\beta J\ast\phi(x,t)),
\end{equation}
where we have introduced the notation $\dot{\phi}(x,t):=\frac{d}{dt}\phi(x,t)$ and for
$b$ we explicit the dependence on $\phi$. Later, when this dependence is not
relevant we will only use $b$.
Thus, such deviating profiles can be viewed as solutions of the
following forced equation:
\begin{equation}\label{forcedeqn}
\frac{d}{dt}m=-m+\tanh(\beta J\ast m)+b,
\qquad
m(x,0)  =  m_0(x),
\end{equation}
where the force term $b$ is some prescribed function of $x$ and $t$.
In this paper, we are interested in investigating the response of the system when
imposing a mean velocity $V$ to the front, i.e., we want to displace the
interface from an initial position $0$ to a final one, $R$,
within a fixed time $T=R/V$.
We consider two scales: the mesoscopic where the interface is diffuse and the
macroscopic where the interface has a sharp jump, i.e., it is given by the step
function $m_\beta(\mathbf 1_{x\geq 0}-\mathbf 1_{x<0})$. 
Let $[0,T]\times\mathbb R$ be the macroscopic time-space domain.
After rescaling back to the mesoscopic variables we are interested in profiles
in the set $\mathcal{U}[\epsilon^{-1}R,\epsilon^{-2}T]$
where
\begin{equation}\label{set}
\mathcal{U}[r,t]=\{
\phi\in C^{\infty}(\mathbb{R}\times(0,t);(-1,1)):\,
\lim_{s\to 0^+}\phi(\cdot,s)=\bar{m},\,
\lim_{s\to t^-}\phi(\cdot,s)=\bar{m}_r
\}
\end{equation}
and where now in the mesoscopic variables the fronts are represented by the instantons
$\bar m$ and $\bar m_r$.
Due to the stationarity of $\bar{m}$, no element in $\mathcal{U}[\epsilon^{-1}R,\epsilon^{-2}T]$
is a solution to the equation \eqref{eqn}. 
Instead, to each element in $\mathcal{U}[\epsilon^{-1}R,\epsilon^{-2}T]$ it corresponds an external force $b$
as in \eqref{functionb},
and in order to select among such forces
one needs to introduce an appropriate action functional.
In \cite{ddp}, the authors invoking linear response theory suggested
the cost functional to be given by $\int_0^{\epsilon^{-2}T}\int_{\mathbb R} b(x,t)^{2}dx\,dt$.
In a companion paper, \cite{BDT}, instead of postulating the cost, we derive it
directly from the underlying stochastic mechanism
via large deviations
over a certain class of functions. 
More precisely, to derive the cost from the stochastic dynamics we work in the
space domain $[-\epsilon^{-1}L,\epsilon^{-1}L]\subset\mathbb R$
with Neumann boundary conditions.
As it will be shown later, the main objects  to which the cost concentrates are the instantons,
which decay exponentially fast as $x\to\pm\infty$ and are well approximated by their finite
volume counterparts as in \eqref{finitevol}.
Hence, in order to avoid unnecessary technical complications we can concentrate here 
in the whole $\mathbb R$
and denote the new cost on $\mathbb R\times [0,\epsilon^{-2}T]$ by:
\begin{equation}\label{newcost}
I_{[0,\epsilon^{-2}T]}(\phi)
=\int_0^{\epsilon^{-2}T}\int_{\mathbb R} \mathcal H(\phi,\dot{\phi})(x,t)\,dx\,dt,
\end{equation}
where for notational simplicity we neglect the dependence of the cost on $\mathbb R$.
The density $\mathcal H(\phi,\dot{\phi})$ is given below and we will also denote it 
by $\mathcal H(x,t)$ in case we do not need to explicit the dependence on $\phi$.
Given $(\phi,\dot{\phi})$ we define
\begin{eqnarray*}
&& u:=\phi\\
&& w:=-\tanh(\beta J\ast\phi)\\
&& b:=\dot{\phi}+\phi-\tanh(\beta J\ast\phi)
\end{eqnarray*}
and after a simple manipulation by a small  abuse of notation we can write $\mathcal{H}$ 
as depending on $(b,u,w)$ in the following form:
\begin{eqnarray}\label{mathcalH}
\mathcal{H}(b,u,w) & = & \frac{1}{2}\left\{
(b-u-w)\log\frac{b-u-w+\sqrt{(b-u-w)^2+(1-u^2)(1-w^2)}}{(1-u)(1-w)}\right.\nonumber\\
& & \mbox{} \left.
-\sqrt{(b-u-w)^2+(1-u^2)(1-w^2)}+1+u w
\right\}.
\end{eqnarray}

The new functional, has a more complicated structure, but asymptotically
has a similar behaviour:
It is a straightforward calculation to see that uniformly on $u\in[-1,1]$ and $w\in(-1,1)$ we have:
\begin{equation}\label{asymptotics}
\lim_{|b|\to\infty}\frac{\mathcal{H}(b,u,w)}{|b|\log(|b|+1)}=
\frac{1}{2}\qquad\mathrm{and}
\qquad
\lim_{|b|\to 0}\frac{\mathcal{H}(b,u,w)}{b^2}=
\frac{1}{4(1+uw)}.
\end{equation}

Note that the cost assumed in \cite{ddp} is approximating the case when $b$ is small, 
but when $b$ is large they are far from each other; hence it
gives a stronger penalization of the deviating profiles than the one derived
from the microscopic system.
As we shall also see in the sequel, the minimizers will correspond to external
fields $b$ which are $\epsilon$-small,
so it is expected that the minimizers of the new functional will be the same with \cite{ddp}. 
But still, we can not exclude a priori the cases that correspond to large external fields
and this is a technical difficulty we have to overcome.
Furthermore, we have a slightly different equation and a more complicated
form of the cost.
Thus, in this paper, we find the minimizer of the derived cost $I_{[0,\epsilon^{-2}T]}(\phi)$ given in \eqref{newcost}
over the class \eqref{set} following the strategy in \cite{ddp} and adjusting the proof
accordingly in order to overcome the aforementioned technical issues. 
To start with, we observe that
the cost of a moving instanton
with $\epsilon$-small velocity, i.e.,
\[
\phi_{\epsilon}(x,t)=\bar{m}_{\epsilon V t}(x),\,\,\,\,\,\,V=\frac{R}{T},
\]
is given by
\[
I_{[0,\epsilon^{-2}T]}(\phi_{\epsilon})
=\frac{1}{4}\|\bar{m}'\|^2_{L^2(d\nu)}V^2 T,
\]
where $\bar{m}'$ is the derivative of $\bar{m}$ and $\|\cdot\|_{L^2(d\nu)}$
denotes the $L^2$ norm on $(\mathbb{R},d\nu(x))$ with
$d\nu(x)=\frac{dx}{1-\bar{m}^2(x)}$.
As in \cite{ddp} it can be shown that other ways to move
continuously the instanton are more expensive.

In such systems one can also observe the phenomenon of nucleations,
namely the appearance of droplets of a phase inside another.
In \cite{BDP} and \cite{BDP2} it has been proved that for such a profile the cost is
bounded by twice the free energy computed at the instanton:

\begin{thm}  
  \label{nucl}
For any $\vartheta>0$ there is $\tau>0$ and a function $\tilde
m_{\epsilon,\tau}(x,s)$, $x\in \mathbb R$, $s\in [0,\tau\epsilon^{-3/2}]$,
symmetric in $x$ for each $s$ and such that
     \begin{equation}
    \label{n2.7}
 \tilde m_{\epsilon,\tau}(x,0) = m_\beta, \quad  \tilde
m_{\epsilon,\tau}(x,\tau\epsilon^{-3/2})=\bar m_{\ell_\epsilon/2}(x), \; x\ge
0,
     \end{equation}
where $\dis{e^{-\alpha \ell_\epsilon}= \epsilon^{3/2}}$, $\alpha>0$ as in
\eqref{1.222}, and
     \begin{equation}
    \label{n2.8}
 I_{\tau\epsilon^{-3/2}}(\tilde m_{\epsilon,\tau})  \le  2 \mathcal
F(\bar m)+ \vartheta.
     \end{equation}
   \end{thm}
Thus, if $V$ gets large, there is a competition between the two values 
of the cost. Therefore, by creating more fronts we can make them move with
smaller velocity with the gain in cost being larger than the extra penalty for
the nucleations.
Following \cite{ddp} we define:     
\begin{equation}
    \label{n2.10}
w_n(R,T):= n  2\mathcal F(\bar m) + (2n+1) \left\{\frac{1}{\mu}
\left(\frac{V}{2n+1}\right)^2 T\right\},
     \end{equation} 
where $\mu=:4\|\bar{m}'\|_{L^2(d\nu)}$ is the mobility coefficient.
The first term is the cost of $n$ nucleations while the second
is the cost of displacement of $2n+1$ fronts (with the smaller velocity $V/(2n+1)$). 
Our main result is given below:

\begin{thm}\label{t2}
Let $P>\inf_{n\geq 0} w_n(R,T)$.
\begin{enumerate}
\item[(i)]
Then $\forall\gamma>0$ 
and for all sequences $\phi_{\epsilon}\in \mathcal{U}[\epsilon^{-1}R,\epsilon^{-2}T]$ with
\begin{equation}\label{finiteness}
I_{\Lambda_\epsilon\times\mathcal T_\epsilon}(\phi_{\epsilon})\leq P,
\end{equation}
we have:
\begin{equation}\label{tobeproved}
\liminf_{\epsilon\to 0}I_{\Lambda_\epsilon\times\mathcal T_\epsilon}(\phi_{\epsilon})\geq \inf_{n\geq 0} w_n(R,T)-\gamma,
\end{equation}
where
$w_n(R,T)$ is given in \eqref{n2.10}.
\item[(ii)]
There exists a sequence 
$\phi_{\epsilon}\in \mathcal{U}[\epsilon^{-1}R,\epsilon^{-2}T]$ such that 
\begin{equation}\label{upperbound}
\limsup_{\epsilon\to 0}
I_{\Lambda_\epsilon\times\mathcal T_\epsilon}(\phi_{\epsilon})\leq 
\inf_{n\geq 0} w_n(R,T).
\end{equation}
\end{enumerate}
\end{thm}

We split the proof in the following sections:
in Section~\ref{prel} we first recall the notions of contours that allow us to separate the phases.
Then we present the multi-instanton manifold and its properties.
This is a repetition of \cite{ddp} and the reader familiar with it could skip it. However,
for completeness of the presentation we also include it here as we will need several of these concepts
in the next sections.
One of the key estimates in the proof is the fact that, because of the finite cost, 
the profiles can not be
away from local equilibrium (instanton manifold) for too long as there is a driving 
gradient force pushing them back.
The main ingredients for this are given in Section~\ref{awayequilibrium} and the key
Proposition~\ref{change} is a bit different than \cite{ddp}, so its proof is adjusted
to the new context.
In Section~\ref{goodandbad} we outline the proof which consists in splitting the time into good/bad time intervals during which the cost is small/large, respectively.
Moreover, we establish the fact that we cannot stay away from the instanton manifold for too long as the 
gradient dynamics drive us back.
Hence, in good time intervals we will eventually find ourselves close to the instanton manifold and, once this happens, we stay there for the whole interval. Then, we can
linearize around some instanton and attribute some velocity to each interface.
This is presented in Section~\ref{linearization}.
Furthermore, we still need to ``connect'' the good time intervals between them and this will be explained in
Section~\ref{fromgoodtogood}.
On the other hand, during bad time intervals which are treated in Section~\ref{badtimes}, 
more interesting things can happen,
namely creation of new fronts (nucleations).
But due to the fact that the overall cost is finite,
they
cannot be too many and the overall
displacement during the bad time intervals is negligible.
Concluding, having split the cost into smooth displacement (with some velocity) and nucleations, we introduce a simplified, closer to macroscopic, model for the motion of the ``centers'' of the instantons.
We call it ``particle model'' and analyze it in Section~\ref{particle} concluding the proof of Theorem~\ref{t2}. Some further technical issues are left for the Appendix.

\section{Preliminaries}\label{prel}

In this section we recall some facts that we will use in the sequel. For a more complete exposition
we refer the reader to the original paper \cite{ddp} and to the monograph \cite{errico}.
We start with the definition of contours and the Peierls estimates which are bounds
on the spatial location of deviations from the equilibrium in terms of the energy $\mathcal{F}$.

\subsection{Contours}

Given $\ell>0$, we
denote by $\mathcal D^{(\ell)}$ the partition of $\mathbb{R}$ into
the intervals $[n\ell,(n+1)\ell)$, $n\in \mathbb Z$, and by
$Q_x^ {(\ell)},\ x\in \mathbb R$ the interval containing $x$ (note
that $x$ need not be the center of $Q_x^ {(\ell)}$). 
We say that $Q_x^ {(\ell)},\ Q_{x'}^ {(\ell)}$ are connected, if
the closures have nonempty intersection, i.e.
$\overline{Q_x^ {(\ell)}}\cap\overline{\ Q_{x'}^ {(\ell)}}\not= \emptyset.$
Now we define
      \begin{equation}
    \label{z5.2}
 m^ {(\ell)}(x):= \frac{1}{|Q_x^ {(\ell)}|}\int_{Q_x^{(\ell)}} m(y)\,dy.
     \end{equation}
Given an ``accuracy parameter''
$\zeta>0$, we introduce
      \begin{equation}
    \label{z5.3}
 \eta^{(\zeta,\ell)}(m;x) =
    \begin{cases} \pm 1 & \text{if $|m^
 {(\ell)}(x)\mp m_\beta|\le \zeta$,}\\0 & \text{otherwise}.
    \end{cases}
     \end{equation}
For any  $\Lambda\subseteq \mathbb R$ which is $D^{(\ell)}$-measurable we call
\begin{eqnarray*}
{\mathcal B}_0^{(\zeta,\ell,\Lambda)}(m)&:=&\left\{x\in \Lambda:\ 
\eta^{(\zeta,\ell)}(m;x)=0\ \right\}\\ 
{\mathcal B}_{\pm}^{(\zeta,\ell,\Lambda)}(m)&:=&
\left\{x\in \Lambda:\ \eta^{(\zeta,\ell)}(m;x)=\pm 1,
\text{ there exists } x'\in \Lambda: \  
\overline{Q_x^ {(\ell)}}\cap\overline{\ Q_{x'}^ {(\ell)}}\not= 
\emptyset
 \ \right.\\  && 
\left. \eta^{(\zeta,\ell)}(m;x')=-\eta^{(\zeta,\ell)}(m;x)
 \right\},\\
{\mathcal B}^{(\zeta,\ell,\Lambda)}(m)&:=&
{\mathcal B}_+^{(\zeta,\ell,\Lambda)}(m)\cup
{\mathcal B}_-^{(\zeta,\ell,\Lambda)}(m)
\cup{\mathcal B}_0^{(\zeta,\ell,\Lambda)}(m).
\end{eqnarray*}

Calling  $\ell_-$ and  $\ell_+$  two  values of the parameter
$\ell$, with $\ell_+$   an  integer multiple of
$\ell_-$,  we define a ``phase indicator''
      \begin{equation*}
\vartheta^{(\zeta,\ell_-,\ell_+)}(m;x) =
    \begin{cases} \pm 1 & \text{if $\eta^{(\zeta,\ell_-)}(m;\cdot)=\pm
    1$ in  $ \left(Q^{(\ell_+)}_{x- \ell_+}
     \cup
Q^{(\ell_+)}_x \cup Q^{(\ell_+)}_{x+  \ell_+}\right), $}
 \\0 & \text{otherwise,}
    \end{cases}
     \end{equation*}
and call contours of $m$ the connected components of the set $\{x:
\vartheta^{(\zeta,\ell_-,\ell_+)}(m;x)=0\}$.  The interval $\Gamma=[x_-,x_+)$ is a
plus contour if $\eta^{(\zeta,\ell_-)}(m;x_{\pm})=1$, a minus
contour if $\eta^{(\zeta,\ell_-)}(m;x_{\pm})=-1$, otherwise it is
called mixed.

Moreover, for any measurable $\Lambda \subseteq\mathbb R$
and $m\in L^\infty(\mathbb R\to[-1,1])$, we define
a local notion of energy by 
\begin{eqnarray*}
{\mathcal F}(m_\Lambda|m_{\Lambda^c})&:=&
\int_\Lambda\phi_\beta(x)d x +\frac{1}{4}
\int_{\Lambda\times \Lambda }
J(x,y)(m(x)-m(y))^2 d y\,d x\\ && +\frac{1}{2}
\int_{\Lambda\times \Lambda^c }
J(x,y)(m(x)-m(y))^2 dy\,d x .
\end{eqnarray*}

The parameters $(\zeta,\ell_-,\ell_+)$ are called {\em compatible} with
$(\zeta_0,c_1,\kappa)\in \mathbb R_+^3$  if $\zeta\in
(0,\zeta_0)$, $\ell_-\le \kappa \zeta$, $\ell_+ \ge 1/\ell_-,$ and
if for any $D^{(\ell_-))}$-measurable set $\Lambda$ and any
$m\in L^\infty(\mathbb R\to[-1,1])$
$$
{\mathcal F}(m_\Lambda|m_{\Lambda^c})
\ge c_1\zeta^2|{\mathcal B}^{(\zeta,\ell_-,\Lambda)}(m)|.
$$
With the above definitions we have:

  \begin{thm}  [\cite{BDP}]
  \label{thmz5.1}
There are positive constants $\zeta_0$, $c_1$, $\kappa$, $c_2$,
so that if  $(\zeta,\ell_-,\ell_+)$ is compatible
with $(\zeta_0, c_1, \kappa)$, then for all $m\in
L^\infty([-L,L];[-1,1])$,
      \begin{equation}
    \label{z5.4}
\mathcal F(m) \ge  \sum_{\Gamma ~{\rm contour~ of~ } m}
w_{\zeta,\ell_-,\ell_+}(\Gamma),
    \end{equation}
where
  \begin{itemize}
\item[] $\dis{
w_{\zeta,\ell_-,\ell_+}(\Gamma)= c_1 \zeta^2 \frac{\ell_-}{\ell_+}
|\Gamma|}$, if $\Gamma$ is a plus or a minus contour;

\item[] $w_{\zeta,\ell_-,\ell_+}(\Gamma) = \max\Big\{ c_1 \zeta^2
\frac{\ell_-}{\ell_+} |\Gamma|\;;\;\mathcal F  (\bar m)-c_2
e^{-\alpha
\ell_+}\Big\}$, if $\Gamma$ is a mixed contour
     \end{itemize}
      and $\alpha$ is given in \eqref{1.222}.
   \end{thm}
From \cite{comets} we have that:
\begin{equation}\label{revcom}
I_{[t_0,t_1]}(\phi)\geq \frac{\beta}{2}
(\mathcal{F}(\phi(\cdot ,t_1))-\mathcal{F}(\phi(\cdot, t_0)))
+\int_{t_0}^{t_1}\|1\wedge|f(\phi)|\|^2_2\, dt.
\end{equation}
Formulas \eqref{revcom}
and \eqref{finiteness} yield
\begin{equation}\label{revertwo}
\sup_{t\leq\epsilon^{-2}T}(\mathcal{F}(\phi_{\epsilon}(\cdot,t))-\mathcal{F}(\phi_{\epsilon}(\cdot,0)))
\leq P,
\end{equation} for every $\phi_{\epsilon}$ in $\mathcal{U}[\epsilon^{-1}R,\epsilon^{-2}T]$. Then, by Theorem \ref{thmz5.1}, for $\zeta$ small enough,
      \begin{equation}
    \label{z5.6}
\sum_{\text{$\Gamma_i$ contours of $u(\cdot,t)$}}\;\;\; |\Gamma_i|\; \le
\; \frac{\ell_+}{c_1\ell_-}  \zeta^{-2} (P+F(\bar m))
    \end{equation}
          \begin{equation}
    \label{z5.7}
\text{number of contours of $u(\cdot,t)$} \le \frac{1}{c_1\ell_-}
\zeta^{-2} (P +F(\bar m))=:N_{\max}
    \end{equation}
          \begin{equation}
    \label{z5.8}
 \text{number of mixed of contours of $u(\cdot,t)$} \le  
\frac{P+F(\bar m)}{\mathcal
 F(\bar m)-c_2 e^{-\alpha
\ell_+}}=: N_{\max}^{\rm mix}
    \end{equation}

\subsection{Multi-instanton  manifold}
     \label{sec:2.1}
The instanton  manifold is the set $\mathcal M^{(1)}=\{\bar m_\xi,
\xi \in \mathbb R\}$.  We extend the notion to the case of several
coexisting instantons by defining the multi-instanton manifold
$\mathcal M^{(k)}$, $k>1$, as the set of all $\bar m_{\bar \xi}$,
$\bar \xi=(\xi_1,\ldots,\xi_k)\in \mathbb R^k,$ $\xi_1<\ldots<\xi_k,$
 sufficiently apart from each other  such that, setting
$\xi_0:=-\infty,\ \xi_{k+1}:=\infty$, the function
  $$
\bar m_{\bar \xi}(x):=\left\{\begin{array}{ll} \bar
m(x-\xi_j)&{\rm if\ } x\in \left[\frac{\xi_{j-1}+\xi_j}{2},
\frac{\xi_{j+1}+\xi_j}{2}\right]\ {\rm and\ } j\ {\rm odd},\\
\bar m(\xi_j-x)&{\rm if\ } x\in \left[\frac{\xi_{j-1}+\xi_j}{2},
\frac{\xi_{j+1}+\xi_j}{2}\right]\ {\rm and\ } j\ {\rm even},
\end{array}\right.
   $$
has exactly $k$ mixed contours.
We denote
    \begin{equation}
    \label{2.1.3.1}
 \mathcal M = \bigsqcup_{k\ge 1} \mathcal M ^{(k)}.
    \end{equation}

To study ``neighborhoods'' of $\mathcal M$ we introduce the notion
of ``center of $m$'' that we use here in a slightly different
sense than usual:

\begin{defin}\label{centerxi}  Recalling $L^2(d\nu_\xi)$, the point $\xi\in \mathbb R$ is a center of $m$ if
$\xi\in \Gamma$, $\Gamma$ a mixed contour of $m$, and if
    \begin{equation}
    \label{2.1.2}
\big( m- \bar m_\xi,\bar m'_\xi\big)_{L^2(d\nu_{\xi})} = 0,\quad \text{or,
equivalently,}\quad \big( m,\bar m'_\xi\big)_{L^2(d\nu_{\xi})} = 0.
    \end{equation}
$\xi$ is an odd, even, center if $\Gamma$ is a $(-,+)$, respectively $(+,-)$
mixed contour.
\end{defin}

\vskip.5cm

The following theorem holds, see \cite{DOPT5},

\vskip.5cm

  \begin{thm}
  \label{thmleip}   If $\zeta$ (in the definition of contours) is
  small enough the following holds.

$\bullet$\; Each mixed contour $\Gamma$ of $m$ contains a center of
$m$.

$\bullet$\; There is $\delta>0$ so that if for some $\xi$ in a
$(-,+)$ mixed contour $\Gamma$ of $m$ (analogous statement holding in
the $(+,-)$ case), $\dis{ \|\text{\bf 1}_\Gamma(m-\bar m_\xi)\|_{L^2(d\nu_{\xi})}\le
\delta}$, then there is a unique center $\xi_m$ in $\Gamma$ and
    \begin{equation}
    \label{2.1.3}
\int_{\mathbb R} \Big( \{ m- \bar m_{\xi'}\}^2-\{m- \bar
m_{\xi_m}\}^2 \Big)> 0,\quad \text{for all $\xi'\in \Gamma$, $\xi'\ne
\xi_m$}
    \end{equation}
and calling  $v=m-\bar m_{\xi}$, $\dis{N_{v,\xi } =  \frac{(v,\bar
m_\xi')}{(\bar m',\bar m')}}$,
 \begin{equation}
    \label{p6.4.1.4}
 \big|\xi_m-(\xi-N_{v,\xi})\big|  \le c \| v\|_{L^2(d\nu_{\xi})}^2, \qquad
|N_{v,\xi}| \le c \| v\|_{L^2(d\nu_{\xi})}.
  \end{equation}
$\bullet$\; If also $\dis{\inf_{\xi' } 
\|{\bf 1}_\Gamma(n-\bar m_{\xi'})\|_{L^2(d\nu_{\xi'})}\le
\delta}$ and $\|m-n\|_{L^2(d\nu_{\xi})}$ is small,
then
     \begin{equation}
     \label{p6.4.1.3.1}
|\xi_m -\xi_n|\le  c \|m-n\|_{L^2(d\nu_{\xi})}.
     \end{equation}
     
  \end{thm}

In Appendix~\ref{L1bound} we will prove the third statement for both the $L^1$ and the $L^{2}$ norm.
By the first statement in Theorem \ref{thmleip} a function $m$
with $k$ mixed contours $\Gamma_1,..,\Gamma_k$ has (at least) one center in
each one of the mixed contours; we denote by $\Xi$ the collection
of all  $\bar \xi=(\xi_1,..,\xi_k)$, $\xi_i<\xi_{i+1}$, $\xi_i$ a
center  of $m$ in $\Gamma_i$ and define
    \begin{equation}
    \label{2.1.4}
d_{\mathcal M}(m) = \inf _{ \bar \xi \in \Xi} \;\;\|m-\bar
m_{{\bar \xi}}\|_{L^2(d\nu_{\bar\xi})}.
    \end{equation}
If $m$ is close enough to $\mathcal M^{(k)}$, then the choice of
$\bar \xi$ is unique.   Note that this definition differs slightly
from the usual definition of a distance of a point from a
manifold, but the following lemma bounds this difference
by replacing the $\inf$ over centers in \eqref{2.1.4}, by the $\inf$ over any generic
variable $\bar \xi \in 
\Gamma_1\times..\times \Gamma_k$, with $\bar\xi=(\xi_1,\ldots,\xi_k)$:

     \begin{lemma}
     \label{lemma0}
For all $k\in\mathbb N$ there are $\delta>0$ and $c>0$ so that if $m$ has $k$
mixed contours $\Gamma_1,..,\Gamma_k$ and $d_{\mathcal M}( m) \le \delta$,
then
    \begin{equation}
    \label{2.1.5}
 d_{\mathcal M}^2( m)
\;\;\ge\;\;\inf_{\bar \xi \in 
\Gamma_1\times..\times \Gamma_k}\;\; \|m-\bar m_{{\bar
\xi}}\|^2_{L^2(d\nu_{\xi})} \;\;\ge\;\; d_{\mathcal M}^2( m) - c \sum_{i=1}^{ k-1}
e^{-\alpha\; \text{\rm dist}(\Gamma_{i+1},\Gamma_i) /2},
    \end{equation}
where $\alpha>0$ is defined  in \eqref{1.222}.
     \end{lemma}

For the proof we refer to \cite{ddp}.

\section{Permanence away from equilibrium}\label{awayequilibrium}

In this section we get bounds on the time interval when a profile
is away from the multi-istanton manifold.
This is done by obtaining a lower bound on the energy gradient
in terms of the distance from the manifold and we will use it in Theorem~\ref{thmm1.1}
in order to get a bound on the number of time intervals
where the given profile is away from local equilibrium.
The main theorem is:

 \begin{thm}
      \label{thmn4.1}
For any   $\vartheta>0$  there is $\rho>0$ such that the following
holds.  Let $m\in  L^\infty(\mathbb R;(-1,1))$ have an odd number
$p$ of mixed contours, let $\mathcal F(m)\le P$ ($P$ as in
Theorem~\ref{t2}) and let $d_{\mathcal M}(m)^2 \ge
\vartheta$. Then
        \begin{equation}
    \label{n4.0}
 \int_{\mathbb R}(1\wedge |f(m)|)^2 \ge \rho,
    \end{equation}
where $f$ is defined in \eqref{gradient}.
     \end{thm}

The proof is essentially contained in \cite{ddp}.
Here we only present the necessary modifications needed for the new functional.
This theorem implies a penalization of the time away
local equilibrium which is stated in the following corollary:

\begin{coro}
    \label{maincor}
Let $\phi$ satisfy \eqref{finiteness}, then for any $\vartheta>0$ there is
$c_{\ref{maincor}}>0$ and $\rho>0$ so that, if $d_{\mathcal M}(\phi(\cdot,t)) \ge \vartheta$
when $t\in [t_0,t_1]$, $0\le t_0<t_1\le \epsilon^{-2}T$, then
necessarily $t_1-t_0 \le \frac{3 P}{c_{\ref{maincor}}\rho}$.
    \end{coro}

{\it Proof.}
By recalling \eqref{revertwo} and from Theorem~\ref{thmn4.1} we obtain that for some $c_{\ref{maincor}}>0$
\[
3P\geq c_{\ref{maincor}}\int_{t_0}^{t_1}\|1\wedge|f(\phi)|\|_2^2\, dt\geq c_{\ref{maincor}}\rho\, (t_1-t_0),
\]
which concludes the proof.
\qed

\medskip

Now we argue as in \cite{ddp}.
We start with the analysis of the condition $d_{\mathcal{M}}(m)^2\geq \vartheta$
when the deviation of $m$ from $\bar{m}_{\bar{\xi}}$
is localized in a neighborhoud of the contours.
We first give the necessary notation.
Let $Q$, $Q_j$ and $B^{\pm}_{k,j}$ be  intervals of the form
$Q=[a,b)$, $Q_j=[a-j,b+j)$, $B^{-}_{k,j}= [a-j-k,a-j)$,
$B^{+}_{k,j}= [b+j,b+j+k)$ with $a,b,j,k$ all in $\ell_+\mathbb
N$. Then, given $\vartheta>0$, we set
\begin{equation}\label{n4.1}
U_{Q,j,\vartheta}=\Big\{ m \in L^\infty(\mathbb R,(-1,1)):\text{ $Q$
is a mixed $\pm$ contour for $m$ and} \,\,
\inf_{\xi\in Q}\int_{Q_j}|m-\bar m_\xi|^2 \geq \vartheta \Big\}
\end{equation}
and
        \begin{equation}
    \label{n4.2}
V_{k,j}=\Big\{ m \in L^\infty(\mathbb R,(-1,1)): \text{
$\eta^{(\zeta,\ell_-)}(m;x)=\pm 1$ for all $x\in
B^{\pm}_{k,j}$}\Big\}.
    \end{equation}

\begin{lemma}
      \label{lemman4.1}
For any $\vartheta>0$, $Q$  and $Q_j$ as above, there is $k$ so
that
        \begin{equation}
    \label{n4.3}
 \int_{Q_{k+j}}|f(m)| >0 \qquad\text{for any  $m\in  
U_{Q,j,\vartheta} \cap
 V_{k,j}$}.
    \end{equation}

     \end{lemma}
The proof is given in \cite{ddp}.
With this lemma we can prove the following:
\begin{prop}
      \label{change}
For any $\vartheta>0$, $Q$  and $Q_j$, let $k$ be as in Lemma
\ref{lemman4.1}.  Then there is $\rho>0$ so that
        \begin{equation}
    \label{n4.5}
\inf_{m \in   U_{Q,j,\vartheta} \cap V_{k,j}} \;\; \int_{Q_{k+j}}
|1\wedge |f(m)||^2 \geq \rho.
    \end{equation}

     \end{prop}

{\it Proof.}
Suppose that the opposite is true. Then there exists a sequence
$m_n\in U_{Q,j,\vartheta} \cap V_{k,j}$
such that
\[
\lim_{n\to\infty}\int_{Q_{k+j}}|1\wedge |f(m_n)||^2=0,
\]
which implies that
$|A_n^c|\to 0$ and $\int_{Q_{k+j}\cap A_n}|f(m_n)|^2\to 0$
where $A_n:=\{x:\,\,|f(m_n(x))|<1\}$. 
We also have that $m_n\rightharpoonup \hat{m}$ in $L^2_{\mathrm{loc}}$
and hence $J\ast m_n\to J\ast\hat{m}$ in $L^2_{\mathrm{loc}}$.
We write (recall that $f(m)=J\ast m-\arctanh m$):
\begin{eqnarray}\label{decompose}
m_n & = & m_n\mathbf 1_{A_n}+m_n\mathbf 1_{A_n^c}=
\tanh(J\ast(m_n\mathbf 1_{A_n})-f(m_n\mathbf 1_{A_n}))\mathbf 1_{A_n}
+m_n\mathbf 1_{A_n^c}\nonumber\\
& = & \tanh(\beta J\ast m_n-f(m_n)\mathbf 1_{A_n})\mathbf 1_{A_n}+m_n\mathbf 1_{A_n^c}.
\end{eqnarray}
Then, $\|m_n\|_{\infty}\leq 1$ implies that
$m_n\mathbf 1_{A_n^c}\to 0$ in $L^2$.
For the first term of $m_n$ in \eqref{decompose} we have:
\begin{eqnarray*}
\int_{Q_{k+j}}|m_n\mathbf 1_{A_n}-\tanh(\beta J\ast\hat{m})|^2
& \leq &
\int_{Q_{k+j}\cap A_n}
|\tanh(\beta J\ast m_n-f(m_n))-\tanh(\beta J\ast\hat{m})|^2\\
& \leq &
c\int_{Q_{k+j}\cap A_n}|f(m_n)|^2\to 0,
\end{eqnarray*}
since $\tanh$ is uniformly Lipschitz continuous.
Thus, $\lim_{n\to\infty} m_n=\tanh(\beta J\ast\hat{m})$ in $L^2(Q_{k+j})$.
Therefore,
since both $m_n\rightharpoonup\hat{m}$ in $L^2_{\mathrm{loc}}$
and $m_n\to\tanh(\beta J\ast\hat{m})$ in $Q_{k+j}$
we obtain that
\[
\hat{m}=\tanh(\beta J\ast\hat{m})\,\,\mathrm{in}\,\, Q_{k+j} 
\,\,\,\,\mathrm{and}
\,\,\, f(\hat{m})(x)=0 \,\,\forall x\in Q_{k+j}.
\]
Now we obtain the contradiction. We have that
\[
\inf_{\xi\in Q}\int_{Q_j}|m_n-\bar{m}_{\xi}|^2\geq\vartheta,\,\,\,\forall n,
\]
which implies (since 
$\lim_{n\to\infty} m_n=\tanh(\beta J\ast\hat{m})$ in $L^2(Q_{k+j})$)
that
\[
\inf_{\xi\in Q}\int_{Q_j}|\tanh(\beta J\ast\hat{m})-\bar{m}_{\xi}|^2\geq\vartheta,
\]
which (since $\hat{m}=\tanh(\beta J\ast\hat{m})$ in $Q_{k+j}$) 
in turn implies that $\hat{m}\in U_{Q,j,\vartheta}$.
Furthermore, $\hat{m}\in V_{k,j}$ (closed in weak $L^2$).
Thus, by lemma~\ref{lemman4.1} there exists $k^*$ such that
$\int_{Q_{k+j}}|f(m)|>0$ for all $m\in U_{Q,j,\vartheta}$.
Contradiction, since this is not true for $\hat{m}$.
\qed

A similar result is true when the external conditions are in the plus or
minus phase. Let
\begin{eqnarray}
 && \hskip-1cm U^{\pm}_{Q,j,\vartheta}
 =\Big\{ m \in L^\infty(\mathbb R,(-1,1)):\; \text{$Q$
is a  $\pm$ contour for $m$ and }  \int_{Q_j}\!\!|m \mp m_\beta|^2
\ge \vartheta \Big\}
   \label{n4.6}
    \end{eqnarray}
        \begin{equation}
    \label{n4.7}
V^{\pm}_{k,j}=\Big\{ m \in L^\infty(\mathbb R,(-1,1)): \text{
$\eta^{(\zeta,\ell_-)}(m;x)=\pm 1$ for all $x\in B^{-}_{k,j}\cup
B^{+}_{k,j}$}\Big\}.
    \end{equation}
Then we also have the following:
\begin{prop}
      \label{propn4.2}
For any $\vartheta>0$, $Q$  and $Q_j$ there are $k$ and $\rho>0$
so that
        \begin{equation}
    \label{n4.8}
\inf_{m \in  U^{ \pm }_{Q,j,\vartheta} \cap V^{ \pm }_{k,j}} \;\;
\int_{Q_{k+j}} (1\wedge |f(m)|)^2 \ge \rho.
    \end{equation}

     \end{prop}

With these ingredients we can conclude the proof of Theorem~\ref{thmn4.1}
following \cite{ddp}.

\section{Strategy of the proof, good and bad time intervals}\label{goodandbad}

Given $\epsilon>0$, we fix an orbit $\phi\in \mathcal{U}[\epsilon^{-1}R,\epsilon^{-2}T]$
as in Theorem~\ref{t2}
(neglecting from the notation the dependence on $\epsilon$)
 and let $b(\phi)$ in \eqref{functionb} be the external force to which it corresponds.
We decompose the time interval $[0,\,\epsilon^{-2}T]$ into subintervals 
$\{S[j,j+1),\ j\in\mathbb{N}\}$ of length $S>0$.
For $\kappa>0$ we choose a parameter 
\begin{equation}\label{delta}
\delta\equiv\delta(\epsilon):=|\log\epsilon|^{-\kappa}
\end{equation}
and define
\begin{equation}
    \label{m1.1}
\phi^{(\delta,S)}(\phi;t) = \begin{cases} 1, &\text{ if
$\dis{\int_{jS}^{(j+1)S}\int_{\mathbb{R}} \mathcal{H}(\phi,\dot{\phi})(x,t) dx\,dt < \delta}$}\\0,
&\text{otherwise}
 \end{cases}\qquad \text{for } t\in S[j,j+1). 
     \end{equation}
To construct ``time contours'' we also define
$\Phi^{(\delta,S)}(\phi;t)$ equal to 1 if
$\phi^{(\delta,S)}(\phi;s)=1$ for all $s\in S[j-1,j+1)$ and $=0$
otherwise. We define $G_{\rm tot}=\{ t \le
\epsilon^{-2}T:\Phi^{(\delta,S)}(\phi;t)=1\}$ and call $t$ a ``good
time'' and $S[j,j+1)$ a ``good time interval" if they are contained in
$G_{\rm tot}$. Bad times and bad intervals are defined complementary.

Given the fact that it is too expensive to be away the instanton manifold (Corollary \ref{maincor}),
the strategy now is to relate the cost functional to the cost of two mechanisms: 
translation of the interfaces and nucleation of new ones.
The first can be achieved by relating the cost to the driving force of the motion of the interface 
and subsequently to its velocity. This is a valid approximation during the ``good'' time intervals.
On the other hand, nucleations can only happen in the ``bad'' ones during which,
the already existing interfaces cannot move too much because the overall cost is finite.
We quantify all this in the next sections. We introduce the velocity of the formed interfaces and relate
it to the cost.
Contrary to \cite{ddp}, for the case of the cost derived via the large deviations this is not
straightforward and new auxiliary profiles have to be introduced.

\subsection{Parameters of the proof.}
We start by choosing some crucial parameters in the estimates.
In Theorem~\ref{nucl} we saw that the cost of a nucleation (producing two fronts)
is close to the cost of creating two interfaces, i.e., close to $2\mathcal F(\bar m)$. 
Since the total cost is bounded by $P$, we obtain an
upper bound ($n^*$) on the total number of fronts:
\begin{equation}\label{nstar}
n^*=1+\frac{2P}{\mathcal F(\bar m)}.
\end{equation}
Moreover, following \cite{ddp}, for given $\gamma>0$ we 
choose a critical value $\ell^{*}$ for the displacement of the fronts, 
after which we consider that a nucleation has occurred.
This is determined to be such that the following holds:
\begin{equation}
    \label{m1.1.0}
\big| \mathcal F\big(\bar m_{(-\ell^*,\ell^*)} \big)- 2 \mathcal
F(\bar m)\big| \le \gamma,
\qquad \mbox{where}\,\,\,\,\,\,\,\bar m_{(-\ell^*,\ell^*)}=\text{\bf 1}_{x\ge 0}\bar m_{ \ell^*
}-\text{\bf 1}_{x< 0}\bar m_{ -\ell^*}.
     \end{equation}
This means that if the profile is made out of a combination of instantons whose 
centers are far enough (more than $2\ell^{*}$) then its free energy is well approximated
by the number of such instantons times the cost of each one of them. Indeed, 
by the  $L^2$-continuity of $\mathcal F(\cdot)$, there is
$\vartheta>0$ so that for all $m$ such that $d_{\mathcal M}(m) \le
\vartheta$ and with centers $(\xi_1,..,\xi_n)$, $n\le n^*$,
$\xi_{i+1}-\xi_i\ge 2\ell^*$, $\forall i$,
we have that:
    \begin{equation}
    \label{m1.1.000}
\big|\mathcal F (m ) -n \mathcal F(\bar m)\big| \le n^*\gamma.
     \end{equation}
However, it may happen that in a newly created nucleation the centers do not exceed the
distance $2\ell^{*}$. 
These are called ``incomplete nucleations'' and we can
neglect them
arguing as in \cite{ddp}, \cite{BDP} and \cite{BDP2} using the propositions below.

We first note that starting with such a profile, the free dynamics make it disappear within
a finite time, depending on the distance $\ell$
(see \cite{BDP}, Proposition 7.1):
\begin{prop}
	 \label{propm1.2}
There is $\tau>0$ so that for any positive $\ell\leq \ell^*$, the solution $v(x,s)$ of \eqref{eqn}
starting from $\bar m_{(-\ell,\ell)}$ (as defined in \eqref{m1.1.0}) verifies 
\begin{equation*}
\sup_{x\in\mathbb R}|v(x,\tau)-m_{\beta}|\leq\vartheta.
\end{equation*}
\end{prop}
This can be also used in a multi-instanton setting:
        \begin{prop}
        \label{propm1.2.1}
There is $L>0$ for which the following holds.  Let $\ell$  and
$\tau$ be as in Proposition \ref{propm1.2} and $\bar
\xi=(\xi_1,...,\xi_n)$,  $n \le n^*$. Call $\mathcal I$ the set of
all even $i$ such that $\xi_{i+1} - \xi_i \le \ell$. Suppose
$\mathcal I$ non void and that for $j\notin \mathcal I$,
$\xi_{j+1} - \xi_j \ge L$. Then the solution $w(x,t)$ of
\eqref{eqn} which starts from $\bar m_{\bar \xi}$ is such that
    \begin{equation}
    \label{m1.1.1}
\sup_{x\in \mathbb R} |w(x,\tau)- \bar m_{{\bar \xi}^*}(x)| \le
 \vartheta,
     \end{equation}
where ${\bar \xi}^*$ is obtained from $\bar \xi$ by dropping all
pairs $\xi_i,\xi_{i+1}$, $i\in \mathcal I$.

     \end{prop}
Then, the same is true if we have an external force whose cost is controlled by a parameter $\alpha>0$.
\begin{prop}
        \label{propm1.3}
Let $\ell$,   $\tau$,  $L$, $\bar \xi$  and ${\bar \xi}^*$ as previously.  
Then there is $\alpha>0$ such that if
    \begin{equation}
    \label{m1.1.3}
    \|m-\bar m_{\bar \xi}\|_2\le \vartheta,\qquad \int_{0}^\tau 
    \int_{\mathbb{R}} |b(x,t)|^2 \,dx\,dt\le \alpha,
        \end{equation}
then  the solution $w(x,t)$ of \eqref{forcedeqn} with force $b$
and which starts from $m$ is such that
    \begin{equation}
    \label{m1.1.4}
\|w(x,\tau)- \bar m_{{\bar \xi}^*}(x)\|_2 \le
4\vartheta.
     \end{equation}

     \end{prop}

\medskip 
From the previous propositions, we fix the parameters $S$ and $\delta$ of our problem. 
Following the analysis in \cite{ddp}
we first choose
the parameter $S$ to be of order one such that:
\begin{equation}\label{S}
S> 10^3\max\big\{\tau,
\frac{3 P}{c_{\ref{maincor}}\rho},\frac 4 \omega
\big\},
       \end{equation}
       where $\omega$ is the spectral gap parameter given in Section~\ref{linearization}.
       On the other hand, for $\delta$ a safe choice would be
      \begin{equation}
      \label{delta-inequality}
\delta = 10^{-3} \min\big\{ \alpha, \frac{\vartheta}{c_{
\ref{prop12.1}}}\big\},\quad \text{$\alpha$ and $c_{
\ref{prop12.1}}$ as in Proposition \ref{propm1.3} and Proposition
\ref{prop12.1}}
       \end{equation}
Hence, our choice in \eqref{delta} satisfies the above criteria.
With this choice of $S$ and $\delta$ we have the following theorem:
\begin{thm}
        \label{thmm1.1}
Let $\phi$ satisfy \eqref{finiteness} and let $\delta$ and $S$ as above.
Then:
    \begin{equation}
    \label{m1.2}
 \text{number of bad time intervals} \le \frac{2 P}{\delta}.
     \end{equation}
If $S[j,j+1)$ is a good time interval,  there is $\dis{t_1\in
S[j-\frac 12 ,j- \frac 14)}$ such that $d_{\mathcal
M}(\phi(\cdot,t_1))\le \vartheta$.
\end{thm}

{\it Proof:}
Suppose that $I$ is a bad interval and let $I^-$ be its previous.
Then inequality \eqref{m1.1} cannot hold for both $I$ and $I^-$ since
otherwise $I$ would have been a good interval.
Hence, the number of bad intervals
is at most twice the number of intervals where \eqref{m1.1} is not true. Thus, 
\[
P>\sum_{I:\,\mathrm{\eqref{m1.1} \,\, is \,\, true}}+\sum_{I:\,\mathrm{\eqref{m1.1}\,\, not \,\, true}}>\frac{1}{2}
(\#\mathrm{bad \,\, intervals})\delta
\]
The second statement follows from Corollary~\ref{maincor}.
\qed

\subsection{Construction of auxiliary profiles $\phi_1$ and $m$.}\label{partition}

Theorem \ref{thmm1.1} allows us to find times $t_j\in[j-\frac 12,j-\frac 14]S,\, 
j\in\mathcal J:=\{1,2,\dots,\frac{\epsilon^{-2}T}{S}\}$
for every good time interval $S[j,j+1)$,
such that
$d_{\mathcal M}(\phi(\cdot,t_j))\le \vartheta$.
Then we define a new partition of $[0, \epsilon^{-2}T]$
as follows: if $S[j,j+1)$ is a good time
interval in the original partition, we replace it by
$[t_j,\, t_{j+1})$ and modify the neighbouring bad time intervals accordingly.
For example, if the previous is bad, in the new partition
it will be replaced by $[S(j-1),\,t_j)$.
If $S[j+1,\, j+2)$ is a good time
interval as well, then $t_{j+1}$ are the ones given by Theorem~\ref{thmm1.1},
 otherwise, $t_{j+1}:=S(j+1)$.
 In this way, we obtain a new, slightly shifted, partition $\{[t_j,t_{j+1})\}_{j\in\mathcal J}$
 of $[0, \epsilon^{-2}T]$.
 Note that in the new partition, the bad time intervals remain unchanged
and this will be relevant in Section~\ref{badtimes}.

\medskip

To prove Theorem~\ref{t2},
we want to derive lower bounds to the cost for a given profile
given the condition on the total displacement.
We estimate the cost of the given profile by
assigning a notion of velocity to its fronts.
The total displacement is then related to the motion
of these fronts with the assigned velocity.
We implement these during the good time intervals.

Suppose $t_j$
is the left endpoint of a maximal connected component
$G$ of $G_{\rm tot}$.
By the definition of $t_j$ we have that
$d_{\mathcal M}(\phi(\cdot,t_j))\le \vartheta$.
For $\vartheta$ small enough, $\phi$ has only mixed contours
which we denote by
$\{\Gamma_i\}_{i=1}^k$, for some $k$ odd. 
We call $\bar{\xi}=(\xi_1,\ldots,\xi_k)$ its centers, ordered increasingly.
In the first good time interval $[t_j,\, t_{j+1})$
of the connected component $G$,
we construct an approximate (to $\phi$) profile $\phi_1$ as well
as another orbit $m$
as follows:
First we truncate the forcing term $b(\phi)$.
For $\lambda>0$ we choose a threshold 
\begin{equation}\label{Delta}
\Delta\equiv\Delta(\epsilon):=|\log\epsilon|^{-\lambda}, \qquad \lambda<\kappa,
\end{equation}
for $\kappa>0$ as in \eqref{delta},
and define a new external field
\begin{equation}\label{bone}
b_1(x,t):=b(\phi)(x,t)\mathbf{1}_{\{(x,t):\,|b(\phi)(x,t)|\leq \Delta(\epsilon)\}}.
\end{equation}
Then we define the auxiliary profiles $\phi_1$ and $m$ to be the solutions of the following system:
\begin{eqnarray}\label{eqnphi1}
\frac{d}{dt}{\phi_1}=-\phi_1+\tanh(\beta J\ast\phi_1)+\alpha b_1, \qquad
\phi_1(\cdot ,t_{\rm in}^+)=\phi(\cdot,t_{\rm in}^+),
\end{eqnarray}
where
\begin{equation}\label{alpha}
\alpha(x,t):=\left(
\frac{1-\bar{m}_{\tilde{\xi}(t)}^2}{8}
\right)^{1/2}.
\end{equation}
The approximate centers $\tilde\xi(t)$, defined in \eqref{center},
are the centers of
the profile $m$
that satisfies the equation:
\begin{eqnarray}\label{eqnm}
\frac{d}{dt}m=-m+\tanh(\beta J\ast m)+b(\phi_1), \qquad
 m(\cdot, t_{\rm in}^+)=m^{\rm in}(\cdot).
\end{eqnarray}
Recall the definition of function $b$ given in \eqref{functionb}. The  time $t_{\rm in}$ and the initial condition $m^{\rm in}(\cdot)$ are given below. For simplicity of the notation
we drop in $t_{\rm in}$ the dependence on $j$.
Note that for the coefficient $\alpha(x,t)$ defined in \eqref{alpha} there exists a large constant $c_*>0$ such that
\begin{equation}\label{cstar}
\frac{1}{c_*}\leq \alpha(x,t)\leq 1, \quad\forall x,t.
\end{equation}
Existence and uniqueness of solutions of the system 
\eqref{eqnphi1}-\eqref{eqnm}
is proved in Appendix~\ref{ex_uniq}.
The idea for introducing the new force $b_1$ is that, following Appendix \ref{bsquare}, for forces of order $\Delta(\epsilon)$,
the density $\mathcal H$ of the cost is well approximated
by $b^2$. 
Moreover, an extra factor $\alpha(x,t)$ is needed in order to reconcile the coefficient
of the asymptotics of $\mathcal H$ (see \eqref{asymptotics})
with the space $L^2(\mathbb R,d\nu_{\bar\xi})$ in which we will be working later
for the linearization around a moving instanton.
Hence, the reason of introducing $\phi_1$ is to have a profile
whose centers are in a controlled distance from those of $\phi$ 
and additionally it has an external force which can be estimated by
the cost.
Then we use the idea in \cite{ddp} of constructing sub-solutions (in our case of $\phi_1$ 
rather than of $\phi$) which start from an appropriately ``regularized"  initial profile 
and whose centers are ensured to move
(being sub-solutions) at least as fast as the corresponding of $\phi$.
We denote this profile by $m$ and note that, by a comparison theorem, it holds that
$m(x,t)\leq\phi_1(x,t)$ for $x\in\mathbb{R}$ and 
$t\in [t_j,\,t_{j+1})$.
Next we present the initial condition $m^{\rm in}(\cdot)$ 
by following the initialization procedure described
in \cite{ddp}, Section 10.

\subsection{Initial condition}

We work in the first good time interval $[t_j, t_{j+1})$. Given $m(\cdot,t_j)$ from equation \eqref{mwithphi},
we construct $m^{\rm in}(\cdot)$ as follows.
Let $\bar\xi(m)=(\xi_1(m),\dots,\xi_k(m))$ be
the centers of $m$ at time $t_j$.

\underline{Case 1}: When $\xi_{j+1}(m)-\xi_j(m)>2|\log\epsilon|^2$ for all $j$.
We let 
$t_{\rm in}=t_j$ and $m(\cdot, t_{\rm in}^+)=m(\cdot, t_{\rm in}^-)$.

\underline{Case 2}: When $\xi_{j+1}(m)-\xi_{j}(m)\leq 2|\log\epsilon|^2$ for some $j$ odd.
We erase both centers for those $j$'s 
and we call the new configuration $\bar\xi^{(1)}(m)$, for which it holds that $\bar m_{\bar\xi^{(1)}(m)}\leq\bar m_{\bar\xi(m)}$. 
Then, we look at all even $j$ in $\bar\xi^{(1)}(m)$ such that $2\ell^*\leq\xi_{j+1}(m)-\xi_j(m)\leq 2|\log\epsilon|^2,\,\ell^*$ as in Proposition~\ref{propm1.3} and we move each $\xi_{j}(m),\,\xi_{j+1}(m)$ to $\xi_{j}'(m),\,\xi_{j+1}'(m)$ so that
 \begin{equation*}
 \xi_{j}(m)+\xi_{j+1}(m)=\xi_{j}'(m)+\xi_{j+1}'(m),\;\;\;\;\xi_{j+1}'(m)-\xi_j'(m)=2|\log\epsilon|^2.
 \end{equation*}
We call $\bar\xi^{(2)}(m)$ the new configuration and $\bar\xi^{(3)}(m)$ the one obtained by $\bar\xi^{(2)}(m)$ following the same procedure as to obtain $\bar\xi^{(1)}(m)$. 
In $\bar\xi^{(3)}(m)$ the pairs $\xi_{j}(m),\,\xi_{j+1}(m)$ with $j$ even either satisfy $\xi_{j+1}(m)-\xi_j(m)\geq 2|\log\epsilon|^2$ or $\xi_{j+1}(m)-\xi_j(m)\leq 2\ell^*$.
\underline{Case 2} is when $\xi_{j+1}(m)-\xi_j(m)\geq 2|\log\epsilon|^2$ for all $j$.
Then, we define 
 \begin{equation*}
 \tilde{m}(x,t_j)=\min\{m(x,t_j),\bar m_{\bar\xi_3(m)}\},
 \end{equation*}
$t_{\rm in}=t_j$ and  $m(\cdot, t_{\rm in})=\tilde m(\cdot, t_j)$.

\underline{Case 3}: This case covers all remaining possibilities in the previous case when in 
$\bar\xi^{(3)}(m)$ there is at least a pair $\xi_{j}(m),\,\xi_{j+1}(m)$ with $j$ even satisfying $\xi_{j+1}(m)-\xi_j(m)\leq 2\ell^*$.
In that case, we let $t_{\rm in}=t_j +\tau$, $\tau$ as in Proposition~\ref{propm1.3} and $m(\cdot,t_{\rm in}^+)$ is the solution at time $t_j +\tau$ of \eqref{eqn} starting from $ \tilde{m}(x,t_j)$.
We finally define $m^{\rm in}(\cdot):=m(\cdot,t_{\rm in})$.

If $j=0$ (and hence $t_j=0$), $m(\cdot,0)$ is the instanton $\bar m(\cdot)$, and initialization is not needed.

As a result of this initialization procedure,
we have that for all $\epsilon>0$ small enough,
the centers of $m(\cdot ,t_j)$ have mutual distance $\geq|\log\epsilon|^2$
and $d_{\mathcal M}(m(\cdot ,t_{\rm in}^+))\leq 6\vartheta$.
To prove this, we use Proposition~\ref{propm1.3} with external
force $b:=b(\phi_1)=\alpha b_1$.
In such a case, we have that $\int b^2$ is related to the cost since
we apply it within a good time interval; hence the requirement \eqref{m1.1.3} is satisfied.
In the next section we show that in the good time interval
$[t_j,\, t_{j+1})$ the solution $m(t,\cdot)$ of \eqref{eqnm}
follows closely a moving instanton $\bar{m}_{\bar{\xi}(t)}$,
where $\bar{\xi}(t)$ are the centers of $m(t,\cdot)$.

\section{Linearization around a moving instanton}\label{linearization}

By the constuction in the previous section, we have that 
in the good time interval $[t_j, \, t_{j+1})$ the profile
$m$ solves the equation
\begin{eqnarray}\label{m}
\frac{d}{dt}{m}=-m+\tanh(\beta J\ast m)+b(\phi_1), \qquad
m(\cdot ,t_j)=m^{\mathrm{in}}(\cdot),
\end{eqnarray}
where the initial condition $m^{\mathrm{in}}(\cdot)$ is given 
by the same initialization as in \cite{ddp}, i.e.,
it has an odd number $k$ of mixed contours at mutual
distance $\geq |\log\epsilon|^2$; moreover  
$d_{\mathcal M}(m^{\mathrm{in}}(\cdot))\le 6\vartheta$.

{\bf Choice of parameters.}  
From \cite{DOPT5} we recall that 
there exists $\omega>0$ such that
\begin{equation}\label{}
(v,Lv )_{L^2(d\nu)}\leq-\omega\|v\|_{L^2(d\nu)},
\end{equation}
for every $v\in L^2(d\nu)$ with $(v, \bar m')_{L^2(d\nu)}=0$, where 
$L$ is the linearized operator of the evolution \eqref{eqn}.
This is called 
``spectral gap parameter''  .
Moreover, let 
$c$ be given in \eqref{c} and $\epsilon_1<\frac{\omega}{8 c}$.
Calling $\bar \xi(t)=(\xi_1(t),..,\xi_k(t))$ the centers of
$m(\cdot,t)$, $t \geq t_j$, we define the approximate centers $\tilde \xi(t)=(\tilde\xi_1(t),..,\tilde\xi_k(t))$ and the deviation $u(\cdot,t)$ as follows:
\begin{equation}
            \label{center}
\big(\mathbf 1_{A_{\alpha^*}}\bar m_{\tilde\xi_i(t)}', [m(\cdot,t)-\sigma_i\bar
m_{\tilde\xi_i(t)}]\big)_{L^2(d\nu)}=0, \qquad u(\cdot,t)=m(\cdot,t)-\bar
m_{\tilde\xi(t)},
       \end{equation}
where
\begin{equation}\label{Aalpha}
A_{\alpha^*}:=\bigg\{x\in\mathbb R:\,\int_{t_{j-1}}^{t_{j+1}} b_1^2(x,s)\,ds\leq\alpha^*\bigg\}
\end{equation}
for $\alpha^*$ small enough
and $\sigma_i= 1$ [$\sigma_i=-1$] if $i$ is odd [even] and $\tilde
\xi_i(t)$ in the $i$-th  mixed  contour of $m(\cdot,t)$. 
From the definition of $A_{\alpha^*}$ we also have that 
\begin{equation}\label{comofAalpha}
|A_{\alpha^*}^c|\leq\frac{8}{\alpha^*}\int_{t_{j-1}}^{t_{j+1}}\|\alpha b_1(s)\|^2_{L^2(d\nu)}ds,
\end{equation}
where
\begin{equation*}
d\nu(x):=\frac{1}{1-{\bar m_{\tilde\xi(t)}}^2}dx.
\end{equation*}
Moreover, we call $\Lambda_i(t)$, $i=1,..,k$, the open intervals
$\dis{\frac 12\big(\tilde\xi_{i-1}(t)+\tilde\xi_i(t),
\tilde\xi_{i+1}(t)+\tilde\xi_i(t)\big)}$, with $\tilde\xi_0(t)=-\infty$
and $\tilde\xi_{k+1}(t)=+\infty$. We have the following estimate
\begin{equation}
|\tilde\xi_{i}(t)-\xi_{i}(t)|+\|u(\cdot,t)-\{m(\cdot,t)-\bar m_{\bar{\xi}(t)}\}\|_{L^2(d\nu)}\leq\frac{c}{\alpha^*}\int_{t_{j-1}}
^{t_{j+1}}\|\alpha b_{1}(s)\|^2_{L^2(d\nu)}ds.
\end{equation}

In the next proposition we give upper
bounds for  displacements of centers with $i$ odd and lower
bounds for those with $i$ even.
In the proof, we follow the strategy in \cite{ddp} with the exception of having a different operator and therefore we have to
work in a appropriately weighted space.

     \begin{prop}
     \label{prop12.1}
There is   a constant  $c_{\ref{prop12.1}}>0$, so that for
$\vartheta$ and $\delta$ small enough and for all  $t\in [t_j, t_{j+1}]$,
we have the following bounds:
     \begin{eqnarray}
      \label{12.6}
&& \!\!\!\!\!\|u( \cdot ,t)\|_{L^2(d\nu)}^2 \le  e^{-(t-t_j)\omega} \|u(
\cdot ,t_{\rm in})\|^2_{L^2(d\nu)} +c_{\ref{prop12.1}} SU_j^2,
 \\&&
\!\!\!\!\!\sigma_i
 [\xi_i(t)-\xi_i(t_{\rm in})] \le -\frac{1}{\|\bar m'\|_2^2}\;
\int_{t_{\rm in}}^t(\alpha b_1,\bar m'_{\xi_i(t)})_{L^2(d\nu)}+
 c_{\ref{prop12.1}}
\big[\|u(\cdot,t_{\rm in})\|_{L^2(d\nu)}^2 +U_j^2\big],
       \label{12.7}
     \end{eqnarray}
    where $i=1,..,k$  and
          \begin{eqnarray}
      \label{12.7.1}
&& U_j^2= 
\int_{t_j}^{t_{j+1}}\|\alpha b_1\|^2_{L^2(d\nu)}+ S R_{\max}, \quad
R_{\max}=c_{\ref{prop12.1}} e^{-\alpha  |\log\epsilon |^2/2}.
     \end{eqnarray}

    \end{prop}
Note that $R_{\max}\to 0$ as $\epsilon \to 0$.

\vskip.5cm

{\it Proof.}
 Let
\[
L:\, L^2(\mathbb{R},d\nu)\to L^2(\mathbb{R},d\nu),\ (Lu)(x):=-u(x)+(1-\bar{m}^2_{\tilde\xi(t)})(\beta J\ast u)(x),
\]
where
\[
d\nu(x):=\frac{dx}{1-\bar{m}^2_{\tilde\xi(t)}(x)}.
\]
For $x\in\Lambda_i$, we have
\begin{equation}
             \label{12.8}
\frac{du(x,t)}{dt} = \sigma_i\dot{\tilde\xi}_i(t)\bar m_{\tilde\xi_i(t)}'+ L u(x,t)+
 \tilde{R}(u)+\alpha b_1(x,t),
      \end{equation}
where
\[
\tilde{R}(u):=G''\big(\beta J\ast(\bar m_{\tilde\xi(t)}+(1-\mu_0)\lambda_0 u)\big)\big(\beta J\ast u
\big)^2,\]
with
\[0\leq\lambda_0,\,\mu_0\leq1
\]
and
\[G(x):=\tanh x.
\]
It is an easy calculation to show that 
\begin{equation}\label{c}
\|\tilde{R}(u)\|_{L^1(d\nu)}\leq c\|u\|^2_{L^2(d\nu)}.
\end{equation}
By multiplying \eqref{12.8} by $u(\cdot,t)\mathbf 1_{A_{\alpha^*}}$ and integrating over
space we obtain:
\begin{equation}\label{derivative}
\frac{\rm d}{{\rm d}t}\left(\frac{1}{2}\|u\mathbf 1_{A_{\alpha^*}}\|_{L^2(d\nu)}^2\right)= 
(u\mathbf 1_{A_{\alpha^*}},\,Lu)_{L^2(d\nu)} +
(u\mathbf 1_{A_{\alpha^*}},\,\tilde{R}(u))_{L^2(d\nu)} +
\int_{\mathbb{R}} u\mathbf 1_{A_{\alpha^*}}\alpha\, b_1 d\nu +R(t),
\end{equation}
where
\begin{equation}
R(t)=\sum_{i=1}^k\sigma_i\dot{\tilde\xi}(t)\left(
\int_{\Lambda_i}\bar{m}_{\tilde\xi_i(t)} u\mathbf 1_{A_{\alpha^*}}\,d\nu+
\int_{\Lambda_i}\bar{m}_{\tilde\xi_i(t)}\frac{\bar{m}_{\tilde\xi_i(t)}}{1-\bar{m}^2_{\tilde\xi_i(t)}}u^2\mathbf 1_{A_{\alpha^*}} d\nu
\right).
 \label{12.8.1.0}
     \end{equation} 
By \eqref{comofAalpha},
\[
\big|(u\mathbf 1_{A_{\alpha^*}},Lu)_{L^2(d\nu)}-(u\mathbf 1_{A_{\alpha^*}},L(u\mathbf 1_{A_{\alpha^*}}))_{L^2(d\nu)}\big|\leq \frac{32}{\alpha^*}\int_{t_{j-1}}^{t_{j+1}}\|\alpha b_1(s)\|^2_{L^2(d\nu)}.
\]     
By the spectral gap property, $(u\mathbf 1_{A_{\alpha^*}},Lu \mathbf 1_{A_{\alpha^*}})_{L^2(d\nu)}\leq-\omega\|u\|_{L^2(d\nu)}$
and by using a similar estimate on $\|u\|_{L^{\infty}}$ as in Theorem C.3 of Appendix in \cite{ddp}
in order to bound the second term in \eqref{derivative}, we obtain:
\begin{eqnarray*}
&& \frac{\rm d}{{\rm d}t}\left(\frac{1}{2}\|u\|_{L^2(d\nu)}^2\right) \leq -\omega\|u\mathbf 1_{A_{\alpha^*}}\|_{L^2(d\nu)}
+c(\epsilon_1+c_1\|u\|_{L^2(d\nu)})^{2/3}\|u\mathbf 1_{A_{\alpha^*}}\|_{L^2(d\nu)}\\
&&\;\;\;\;\;\;\;\;\;\;\;\;\;\;\;+(u \mathbf 1_{A_{\alpha^*}},\alpha b_1)_{L^2(d\nu)}+c' \int_{t_j}^{t_{j+1}} \|\alpha b_1(s)\|_{L^2(d\nu)}^2 +R(t).
\end{eqnarray*}
Let
       \begin{equation}
             \label{12.9.0}
\tau:=\inf\left\{t: \|u(\cdot,t)\|_{L^2(d\nu)}^{2/3}>\frac
{\omega}{8 c c_1}\right\}.
       \end{equation}
Bounding $\|(u\mathbf 1_{A_{\alpha^*}},\alpha b_1)\|_{L^2(d\nu)} \le \frac{2\|\alpha b_1\|^2_{L^2(d\nu)}}
{\omega}
+\frac{\omega\|u\mathbf 1_{A_{\alpha^*}}\|^2_{L^2(d\nu)}}{4}$, for all times $t\in [t_j,
t_{j+1}]$ such that $t<\tau$ we have:
      $$
\frac{\rm d}{{\rm d}t}\left(\frac{1}{2}\|u\mathbf 1_{A_{\alpha^*}}\|_{L^2(d\nu)}^2\right)\le
-\frac{\omega}{2}
\|u\mathbf 1_{A_{\alpha^*}}\|_{L^2(d\nu)}^2+\frac{2}{\omega}\|\alpha b_1\|_{L^2(d\nu)}^2+R(t),
       $$
i.e., for $t^*={\rm min}\{\tau, t_{j+1}\}$ we obtain
     $$
\|\mathbf 1_{A_{\alpha^*}}u(\cdot,t^*)\|_{L^2(d\nu)}^2\le e^{-(t^*-t_j)
\omega}\|u(\cdot,t_j)\|^2_{L^2(d\nu)} +c_{\ref{prop12.1}}\left(
\int_{t_j}^{t^*} \|\alpha b_1(s)\|_{L^2(d\nu)}^2+S R_{\max}\right),
      $$
with $R_{\max}$ defined in \eqref{12.7.1}. Since 
\[
\|u\|_{L^2(\rm d \nu)}^2\leq\|\mathbf 1_{A_{\alpha^*}} u\|_{L^2(\rm d \nu)}^2+\frac{4}{\alpha^*}\int_{t_j}^{t_{j+1}} \|\alpha b_1(s)\|_{L^2(d\nu)}^2 ,
\]
we have 
  $$
\|u(\cdot,t^*)\|_{L^2(d\nu)}^2\le e^{-(t^*-t_j)
\omega}\|u(\cdot,t_j)\|^2_{L^2(d\nu)} +c_{\ref{prop12.1}}\left(
\int_{t_j}^{t^*} \|\alpha b_1(s)\|_{L^2(d\nu)}^2+S R_{\max}\right).
      $$

By the choice of $\delta$ in \eqref{delta-inequality} and \eqref{errc0} we have
\[
c_{\ref{prop12.1}}\int_{t_j}^{t^*} \|\alpha b_1(s)\|_{L^2(d\nu)}^2+S R_{\max}
\leq c_{\ref{prop12.1}}\left(\frac{1}{1-c_{*}^{2}C\Delta(\epsilon)}\delta+S R_{\max}\right)\leq 10^{-3}.
\]
Thus, for $\delta$, $\vartheta$ and $\epsilon$ small enough,
$\|u(\cdot,t^*)\|_{L^2(d\nu)}^2\le (\frac{\omega}{8 c c_1})^3$
and hence $t^*=t_{j+1}$.

For the proof of \eqref{12.7}, we multiply
\eqref{12.8} by $\mathbf 1_{A_{\alpha^*}}\bar{m}'_{\tilde\xi_i(t)}$ and estimate
$(\mathbf 1_{A_{\alpha^*}}\bar{m}'_{\tilde\xi_i(t)},u_t)_{L^2(d\nu)}$
by first writing \eqref{center} as
\begin{equation}\label{one}
(\mathbf 1_{A_{\alpha^*}}\bar{m}'_{\tilde\xi_i(t)},
\sigma_i\bar{m}_{\tilde\xi_i(t)}-\bar{m}_{\tilde\xi(t)})_{L^2(d\nu)}
=(\mathbf 1_{A_{\alpha^*}}\bar{m}'_{\tilde\xi(t)},u)_{L^2(d\nu)},
\end{equation}
after adding and
subtracting $\bar{m}_{\tilde\xi(t)}$.
Since the measure $d\nu$ depends on time, we also have:
\begin{eqnarray}
\frac{d}{dt}(\mathbf 1_{A_{\alpha^*}}\bar{m}'_{\tilde\xi_i(t)},u)_{L^2(d\nu)} & = &
(\mathbf 1_{A_{\alpha^*}}\bar{m}'_{\tilde\xi_i(t)},u_t)_{L^2(d\nu)}
+(\mathbf 1_{A_{\alpha^*}}\bar{m}_{\tilde\xi_i}''\sigma_i\dot{\tilde\xi}_i,u)_{L^2(d\nu)}\nonumber\\
&& +\sum_j\int_{\Lambda_j}\bar{m}_{\tilde\xi_i}' \mathbf 1_{A_{\alpha^*}}u
\frac{2\bar{m}_{\tilde\xi_j}\bar{m}_{\tilde\xi_j}'\dot{\tilde\xi}_j}
{(1-\bar{m}_{\tilde\xi_j}^2)^2}dx.
\end{eqnarray}
We obtain:
\begin{eqnarray}\label{A}
(\mathbf 1_{A_{\alpha^*}}\bar{m}'_{\tilde\xi_i},u_t)_{L^2(d\nu)} & = &
\dot{\tilde\xi}_i\left\{
(\mathbf 1_{A_{\alpha^*}}\bar{m}_{\tilde\xi_i}'',u)_{L^2(d\nu)}
+(\mathbf 1_{A_{\alpha^*}}\bar{m}_{\tilde\xi_i}'',\bar{m}_{\tilde\xi}-\sigma_i\bar{m}_{\tilde\xi_i})_{L^2(d\nu)}
\right\}\nonumber\\
&&
-\sum_{j\neq i}\left(
\mathbf 1_{A_{\alpha^*}}\mathbf 1_{\Lambda_j}\bar{m}'_{\tilde\xi_i},
(\sigma_i\dot{\tilde\xi}_i\bar{m}_{\tilde\xi_i}'-\sigma_j\dot{\tilde\xi}_j\bar{m}_{\tilde\xi_j}')
\right)_{L^2(d\nu)}\nonumber\\
&&
+\sum_{j\neq i}\int_{\Lambda_j}2 \mathbf 1_{A_{\alpha^*}}u\bar{m}_{\tilde\xi_j}
\frac{\bar{m}_{\tilde\xi_i}'\bar{m}_{\tilde\xi_j}'}{1-\bar{m}_{\tilde\xi_j}^2}d\nu\nonumber\\
&&
-\sum_{j\neq i}\int_{\Lambda_j}\mathbf 1_{A_{\alpha^*}}\bar{m}_{\tilde\xi_i}'
(\sigma_i\bar{m}_{\tilde\xi_i}-\bar{m}_{\tilde\xi})\frac{1}{1-\bar{m}_{\tilde\xi_j}^2}
2\bar{m}_{\tilde\xi_j}\bar{m}_{\tilde\xi_j}'\sigma_j\dot{\tilde\xi}_j d\nu.
\end{eqnarray}
On the other hand, in \eqref{12.8} we have:
\[
(\mathbf 1_{A_{\alpha^*}}\bar{m}_{\tilde\xi_i}',L u)_{L^2(d\nu)}=(u,L\bar{m}_{\tilde\xi_i}')_{L^2(d\nu)},
\,\,\,\,{\rm with}\,\,\, |L\bar{m}_{\tilde\xi_i}'|\leq R_{\max}.
\]
Thus, from \eqref{12.8} and \eqref{A} we obtain:
\[
\sigma_i\dot{\tilde\xi}_i\left[
\|\bar{m}_{\tilde\xi_i}'\mathbf 1_{A_{\alpha^*}}\|^2_{L^2(d\nu)}-\sigma_i\{(\mathbf 1_{A_{\alpha^*}}\bar{m}_{\tilde\xi_i}'',u)_{L^2(d\nu)}
+(\mathbf 1_{A_{\alpha^*}}\bar{m}_{\tilde\xi_i}'',\bar{m}_{\tilde\xi}-\sigma_i\bar{m}_{\tilde\xi_i})_{L^2(d\nu)}\}
\right]
\]
\[
+\sum_{j\neq i}\left(\mathbf 1_{A_{\alpha^*}}
\mathbf 1_{\Lambda_j}\bar{m}'_{\tilde\xi_i},
(\sigma_i\dot{\tilde\xi}_i\bar{m}_{\tilde\xi_i}'-\sigma_j\dot{\tilde\xi}_j\bar{m}_{\tilde\xi_j}')
\right)_{L^2(d\nu)}
-\sum_{j\neq i}\int_{\Lambda_j}1_{A_{\alpha^*}}2 u\bar{m}_{\tilde\xi_j}
\frac{\bar{m}_{\tilde\xi_i}'\bar{m}_{\xi_j}'}{1-\bar{m}_{\xi_j}^2}d\nu
\]
\[
+\sum_{j\neq i}\int_{\Lambda_j}1_{A_{\alpha^*}}\bar{m}_{\tilde\xi_i}'
(\sigma_i\bar{m}_{\tilde\xi_i}-\bar{m}_{\tilde\xi})\frac{1}{1-\bar{m}_{\tilde\xi_j}^2}
2\bar{m}_{\tilde\xi_j}\bar{m}_{\tilde\xi_j}'\sigma_j\dot{\tilde\xi}_j d\nu
\]
\[
\leq -(\bar{m}_{\tilde\xi_i}',\alpha b_1)_{L^2(d\nu)}+|A_{\alpha^*}^c|+c' c\|\mathbf 1_{A_{\alpha^*}}u\|_{L^2(d\nu)}^2+R_{\max}
\]
which has the form:
\begin{equation}\label{form}
\sigma_i\|\bar{m}'\|_{L^2(d\nu)}^2\dot{\tilde\xi}_i(t)\leq\beta_i+\sum_{j=1}^k
a_{i,j}|\dot{\tilde\xi}_j|,
\end{equation}
where
\begin{equation}\label{beta}
\beta_i=(\mathbf 1_{A_{\alpha^*}}\bar{m}_{\tilde\xi_i}'',u)_{L^2(d\nu)}-(\bar{m}_{\tilde\xi_i}',\alpha b_1)_{L^2(d\nu)}+c' c\|u\|_{L^2(d\nu)}^2
+|A_{\alpha^*}^c|+R_{\max},
\end{equation}
with
\[
|\beta_i+(\bar{m}_{\tilde\xi_i}',\alpha b_1)_{L^2(d\nu)}|\leq
c''[
e^{-(t-t_{\rm in})\omega}\|u(\cdot,t_{\rm in})\|_{L^2(d\nu)}^2\]
\[
\;\;\;\;\;\;\;\;\;\;\;\;\;\;\;\;\;\;\;\;+\int_{t_{\rm in}}^t\|\alpha b_1\|_{L^2(d\nu)}^2 ds+S R_{\max} 
+ \|1-\mathbf 1_{A_{\alpha^*}}\|_{L^2(d\nu)}\|\alpha b_1\|_{L^2(d\nu)}
\]
and
\begin{eqnarray}\label{aij}
a_{i,j} & = & (\mathbf 1_{\Lambda_j}\bar{m}_{\tilde\xi_j}'',\bar{m}_{\tilde\xi_i}-\sigma_j\bar{m}_{\tilde\xi_j})
+(\mathbf 1_{\Lambda_j}\bar{m}_{\tilde\xi_i}',\bar{m}_{\tilde\xi_j}')_{L^2(d\nu)}\nonumber\\
&& +\int_{\Lambda_j}2 u \bar{m}_{\tilde\xi_j}\frac{\bar{m}_{\tilde\xi_i}'\bar{m}_{\tilde\xi_j}'}
{1-\bar{m}_{\tilde\xi_j}^2}d\nu-\int_{\Lambda_j}\bar{m}_{\tilde\xi_i}'
(\sigma_i\bar{m}_{\tilde\xi_i}-\bar{m}_{\tilde\xi})\frac{2\bar{m}_{\tilde\xi_j}\bar{m}_{\tilde\xi_j}'}
{1-\bar{m}_{\tilde\xi_j}^2}d\nu.
\end{eqnarray}
Then we conclude the proof in the same fashion as in \cite{ddp}
by estimating $a_{i,j}$, since $\xi_i$ and $\xi_j$ are well separated.
\qed

\medskip
Concluding this section, we recall that we constructed $m(t,\cdot)$ for $t\in[t_j,\, t_{j+1}]$
and obtained estimates for the error
$\|m(\cdot ,t)-\bar{m}_{\bar{\xi}(t)}\|_{L^2(d\nu)}^2$.
Next we define $m(\cdot,t_{j+1}^+)$ in order to
apply this linearization procedure in the whole of the maximal connected
component $G$.

\section{From a good time interval to the next}\label{fromgoodtogood}

The result of Proposition~\ref{prop12.1} ensures that during
the good time interval $[t_j,t_{j+1})$ the solution of \eqref{m} is close
to a moving instanton.
More precisely, by \eqref{delta-inequality} we have that $ c_{\ref{prop12.1}} U_{j}^2 \le
\vartheta$ and by \eqref{S} that $e^{-\omega S}\le 1/2$. 
Then
by \eqref{12.6} we get, supposing $\epsilon$ small enough,
     \begin{equation}
      \label{12.199}
 \|u(\cdot,t_{j+1})\|_{L^2(d\nu)}^2 \le  e^{-\omega S}\|u(t_j)\|^2_{L^2(d\nu)}
+ c_{\ref{prop12.1}} U_{j}^2   \le 4\vartheta.
     \end{equation}
Furthermore,
since $\xi_{i+1}(t_{j+1})-\xi_{i}(t_{j+1}) \geq
 |\log\epsilon|^2/2$,
as we have seen in the course of the proof of Proposition
\ref{prop12.1}, it follows from  \eqref{2.1.5} that for
$\epsilon$ small enough,
     \begin{equation}
      \label{12.199.0} 
d_{\mathcal M} (m(\cdot,t_{j+1})) \le  5\vartheta.
     \end{equation}
We introduce the notion of {\it velocity of a front} $\bar{m}_{\xi_i}(t)$, by defining:
\begin{equation}
      \label{12.20a}
   v^0_i(t):=\sigma_i\frac{1}{\|\bar m'\|_{L^2(d\nu)}^2}\;
\Big|(\alpha b_1,\bar m'_{\xi_i(t)})_{L^2(d\nu)}\Big|,
    \end{equation}
where again $\sigma_i= 1$ [$\sigma_i=-1$] if $i$ is odd [even].
Moreover, we want to control the position of the centers of $m(\cdot ,t)$,
so we denote by $r_i(t)$ the leftmost [rightmost]
position of the center $\xi_i$ of $m(\cdot ,t)$, for $i$ odd [even], taking into account
the error in determining the position $\xi_i$.
Thus, the position $r_i(t)$ will be given by $\xi_i$ plus the integral of the velocity
induced by the error
$\|m(\cdot ,t)-\bar{m}_{\bar{\xi}(t)}\|_{L^2(d\nu)}^2$.
We define:
    \begin{eqnarray}
           \label{12.20b}
&&   v_i(t):=v_i^0(t)+ \sigma_i c_{\ref{prop12.1}}\Big( U_j^2 + {\|u
(\cdot,t_j)\|^2_{L^2(d\nu)}} \Big),  \\&&
 r_i(t):= \xi_i(t_j)
 +\int_{t_j}^t v_i (s),\qquad \bar r(t)=\big(r_1(t),..,r_k(t)\big).
       \label{12.200}
     \end{eqnarray}
Notice that $\bar{r}(t)\leq\bar{\xi}(t)$ for $t\in [t_j,t_{j+1})$, where the partial order is 
defined as:\begin{equation}\label{partialorder1}
(\xi_1,...,\xi_k)\geq (\xi'_1,...,\xi_{k'}')\;\Leftrightarrow \;\bar m_{(\xi_1,...,\xi_k)}\geq\bar m_{(\xi_1',...,\xi_{k'}')}.
\end{equation}
In particular, if $k=k'$,
\begin{equation}\label{partialorder2}
(\xi_1,...,\xi_k)\geq (\xi'_1,...,\xi_k')\;\Leftrightarrow \;\xi_{i}\leq\xi_{i}', \mathrm{i \;odd,} \;\;\;\xi_{i}\geq\xi_{i}' ,\;\mathrm{i \;even}.
\end{equation}

By the definition of $t_{j+1}$ we know that 
$d_{\mathcal M}(\phi(\cdot,t_{j+1}))\leq \vartheta$.
Suppose now that, for $\epsilon>0$ small enough,
$\phi(\cdot,t_{j+1})$ has $k'$-many mixed contours 
$\{\tilde{\Gamma}_i\}_{i=1,\ldots,k'}$, $k'$ odd,
with 
$\|\mathbf 1_{\tilde{\Gamma}_i}(\phi-\bar{m}_{\tilde{\xi}_i})\|_{L^2}\leq\vartheta$
for some $\tilde{\xi}_i\in\tilde{\Gamma}_i$, $i=1,\ldots,k'$.
Note that in general $k'\neq k$ (since $m$ has been re-initialized at $t_{j}$ and some
fronts might have been cancelled).
Then by Theorem~\ref{thmleip} we have that there exist unique centers
$\{\xi_i(\phi)(t_{j+1})\}_{i=1,\ldots,k}$ of $\phi(\cdot,t_{j+1})$.
The strategy goes as follows:
note that since (using \eqref{cstar})
\[
|b(\phi_1)|=|\alpha b_1|=\bigg|\bigg(\frac{1-\bar m_{\tilde\xi(t)}^2}{8}\bigg)^{1/2}b_1\bigg|\leq |b_1|\leq |b(\phi)|,
\]
the profile $\phi_1(t_{j+1})$ is expected to have 
its odd [even] indexed centers
on the left [right] of the corresponding centers of $\phi(t_{j+1})$.
On the other hand, the profile $m(t_{j+1})$, being a sub-solution of the
equation $b(m)=b(\phi_1)$, with initial condition 
$m(t_{j})$ re-initialized as before, it has its odd [even] centers on the right [left]
of the corresponding centers of $\phi_1(t_{j+1})$.
However, it is not guaranteed that this
is also the case with the centers of $\phi(t_{j+1})$.
Therefore, since in the next good time interval we choose $\phi_1(\cdot,t_{j+1}^+):=\phi(\cdot,t_{j+1}^+)$
we need to re-initialize $m(\cdot,t_{j+1}^+)$ to be such that
$m(\cdot,t_{j+1}^+)\leq\phi_1(\cdot,t_{j+1}^+)$ and keep track of the relevant error.
As a result of the initialization,
the profile $m(t_{j+1})$ may have fewer centers than 
$\phi_1(\cdot,t_{j+1})$.

We estimate the distance between the 
corresponding
centers of $\phi$ and $m$ at $t_{j+1}$, when both are close to the manifold $\mathcal M$.
Recall also that, by the initialization, the centers at $t_j^+$
have mutual distance $\geq|\log\epsilon|^2$.
To perform our estimate we introduce an auxiliary profile $\phi_2$ 
by putting as forcing term only $b_1$ with the same initial condition. 
For $t\in[t_j,t_{j+1})$ we have:
\[
\|\phi(t)-\phi_2(t)\|_{L^1}\leq\int_{t_j}^te^{-(t-s+t_j)}\beta\|J\|_{L^1}
\|\phi(s)-\phi_2(s)\|_{L^1}ds+\int_{t_j}^t\int_{\mathbb{R}}e^{-(t-s+t_j)}|b-b_1|dx\,ds,
\]
where
\[
\int_{t_j}^t \int_{\mathbb{R}}e^{-(t-s+t_j)} |b-b_1|dx\, ds\leq\int_{|b|>\Delta(\epsilon)\}}
 |b| dx\, ds.
\]
In the good time interval $[t_j,t_{j+1})$ we define the quantity:
\begin{equation}\label{deltaj}
\delta_j:=\int_{t_j}^{t_{j+1}}\int_{\mathbb{R}}\mathcal{H}(b,u,w)(x,s)ds\,dx,
\end{equation}
in which case it is of the order $\delta(\epsilon)$.
From \eqref{asymptotics} we obtain that:
\begin{eqnarray*}
\delta_j=\int_{t_j}^{t_{j+1}}\int_{\mathbb{R}}\mathcal{H}(b,u,w)(x,s)ds\,dx
&\geq&
\int_{\{|b|>\Delta(\epsilon)\}}\mathcal{H}(b,u,w)(x,s)ds\, dx \\
&\geq&
C\int_{\{|b|>\Delta(\epsilon)\}}|b|\log(|b|+1)ds\, dx \\
&
\geq &
C \int_{\{|b|>\Delta(\epsilon)\}} |b|\log(1+\Delta(\epsilon))ds\, dx .
\end{eqnarray*}
Thus, (since $\|J\|_{L^1}=1$)
\begin{equation}\label{Ione}
\|\phi(\cdot,t)-\phi_2(\cdot,t)\|_{L^1}\leq \beta\int_{t_j}^te^{-(t-s+t_j)}\|\phi(\cdot, s)-\phi_2(\cdot, s)\|_{L^1}ds+\frac{\delta_j}{C\log(1+\Delta(\epsilon))}
\end{equation}
and for a new constant $C>0$ by Gronwall's lemma we obtain that
\begin{equation}\label{phiandphi2}
\|\phi(\cdot, t_{j+1})-\phi_2(\cdot,t_{j+1})\|_{L^1}\leq C
e^{(2+\beta)S}
\frac{\delta_j}{\Delta(\epsilon)}.
\end{equation}
On the other hand, comparing to $m$ we have
\[
\frac{d}{dt}\int_{\mathbb{R}} (\phi_2-m)^2(x,t)\,dx =
\]
\begin{eqnarray*}
& = & 
-2\int_{\mathbb{R}}(\phi_2-m)^2(x,t)\,dx+2\int_{\mathbb{R}} (1-\alpha)b_1(x,t)(\phi_2-m)(x,t)\,dx\\
&& +2\int_{\mathbb{R}} (\phi_2-m)(x,t)(\tanh(\beta J\ast \phi_2(x,t))-\tanh(\beta J\ast m^0(x,t)))dx\\
&\leq&
C\int_{\mathbb{R}}(\phi_2-m)^2(x,t)\,dx+c\int_{\mathbb{R}} (1-\alpha)^{2} b_1^2(x,t)\,dx.
\end{eqnarray*}
Since from \eqref{cstar} it holds that $1-\alpha\leq (c^{*}-1) \alpha$,
applying Gronwall's inequality and using \eqref{errc0} we obtain
\begin{eqnarray}\label{Itwo}
\|\phi_2(\cdot,t)-m(\cdot,t)\|_{L^2}^2
& \leq &
ce^{(2+\beta)(t-t_{j})} \int_{\mathbb{R}} \int_{t_j}^t\alpha^2 b_1^2ds\,dx \nonumber\\
&\leq&
ce^{(2+\beta)S}\frac{1}{1-c_{*}^{2}C\Delta(\epsilon)}\,\delta_j,
\end{eqnarray}
for $\epsilon$ small enough so that $c_{*}^2C\Delta(\epsilon)<1$.
Thus, since $\|\phi(\cdot, t)-\phi_2(\cdot,t)\|_{L^\infty(\mathbb R)}\leq 2$, \eqref{Ione} and \eqref{Itwo} yield
\begin{eqnarray}\label{jumpS2}
\|\phi(\cdot, t)-m(\cdot, t)\|_{L^2(\mathbb R)}^2 &\leq& 
2\|\phi_2(\cdot,t)-m(\cdot,t)\|_{L^2(\mathbb R)}^2(t)+4\|\phi(\cdot, t)-\phi_2(\cdot,t)\|_{L^1(\mathbb R)}\nonumber\\
&\leq& 
C\frac{\delta_j}{\Delta(\epsilon)}+ce^{(2+\beta)S}\frac{1}{1-c_{*}^{2}C\Delta(\epsilon)}\,\delta_j=:S_{\epsilon}^j,
\end{eqnarray}
where by choosing $\kappa<\lambda$ in the definition of $\Delta(\epsilon)$ in \eqref{Delta}, we have that $S^j_{\epsilon}\to 0$ as $\epsilon\to 0$.
Using the above estimate and the fact that both $m$ and $\phi$ are close to the manifold at time $t_j$, we
obtain that 
\begin{equation}\label{jumpS3}
|\xi(\phi)(t_{j+1})-\xi(m)(t_{j+1})|\leq
\|\bar m_{\bar\xi(m)}-\bar m_{\bar\xi(\phi)}\|\leq S_{\epsilon}^j+6\vartheta.
\end{equation}
Next, recalling the definition of $\bar r(t)$ in \eqref{12.200}, in order to define $r_i(t_{j+1}^+)$
we consider the quantity 
\begin{equation}\label{rhat}
\hat r_i(t_{j+1}):=r_i(t_{j+1})+\sigma_i S^j_{\epsilon}
\end{equation}
and we erase
all pairs $i$, $i+1$ such that
$\hat{r}_{i+1}(t_{j+1})-\hat{r}_i(t_{j+1})\leq|\log\epsilon|^2$.
Then we let
\[
r_i(t_{j+1}^+):=\hat r_i(t_{j+1}),
\]
if no such erasing has occurred for the index $i$. Otherwise, we let
$r_i(t_{j+1}^+):=\emptyset$.

In Section~\ref{particle} we introduce the notion of particles while referring to
the fronts and we say that in this case the particles
$i$ and $i+1$ have collided and, due to this collision, they disappeared.
We will also write that $r_i(t)=r_{i+1}(t)=\emptyset$ for $t>t_{j+1}$.
Moreover, note that the function $\bar{r}(t)$ has jumps at the times between
good time intervals and this fact will be taken into account in the
estimation of the total displacement and the corresponding ``macroscopic" cost
expressed in terms of the cost due to the motion of the particles.
For the re-initialization at $t_{j+1}^+$ we define:
\begin{equation}\label{constrofm}
m(\cdot,t_{j+1}^+):=\min\{
\phi(\cdot,t_{j+1}),\,\bar{m}_{r_i(t_{j+1}^+)}(\cdot)
\}.
\end{equation}
In this way we ensure that $m(\cdot,t_{j+1}^+)\leq\phi(\cdot,t_{j+1})$ as well as 
that $r_i(t_{j+1}^+)$ is a lower [upper]
bound of $\xi_i(m(\cdot,t_{j+1}^+))$ for $i$ odd [even]. 
Thus, taking $\epsilon$ small enough we have that
$d_{\mathcal M}(m(\cdot,t_{j+1}^+))\leq 20\vartheta$
and that its centers have mutual distance $\geq|\log\epsilon|^2$.
So we can repeat the same procedure for the next good time interval
$[t_{j+1},\, t_{j+2})$.

\section{Displacement during the bad time intervals}\label{badtimes}

From \eqref{m1.2}
the maximal length of the connected component of bad time intervals is bounded by 
$
S\frac{P}{\delta(\epsilon)}<<|\log\epsilon|^{2}
$
for the choice of $\delta(\epsilon)$ 
made in \eqref{Delta}. Moreover, the applied force $b$ can be related to and 
bounded by the cost.
Therefore, the displacement of the already existing centers should be smaller than 
$|\log\epsilon|^{2}$, which is the distance between the appropriately initialized
centers of the interfaces.
Similarly, the newly nucleated fronts are also at a distance from each other 
smaller than $|\log\epsilon|^{2}$ even at the end of the connected component of 
the bad time intervals.
Hence, overall
the motion during the bad time intervals will be negligible macroscopically.

Suppose that $[t_{j'},t_{j''})$ is a connected component of bad time intervals. 
Recalling the construction of the partition of good and bad time intervals in
subsection~\ref{partition}, we have that $t_{k}=k S$, for all $j'\leq k \leq j''$, $k\in\mathbb N$.
In the connected component of bad time intervals we define the profile $m$ by solving the
equation
\begin{eqnarray}\label{mwithphi}
\frac{d}{dt}{m}=-m+\tanh(\beta J\ast m)+b(\phi),
\end{eqnarray}
with initial condition the profile $m(t_{j'},\cdot)$ as we obtained it from the previous
good time interval.
Invoking again Corollary \ref{maincor} and the choice of $S$ for the profile $m$ constructed above, for $j'+1\leq k\leq j''$
there exist $\bar t_{k} \in[t_j, t_{j+1})$ with
$m(\bar t_{k},\cdot)$ close to $\mathcal M$.

We compare the solution $m$ to the solution $m^{0}$ of the same equation
without the forcing term $b(\phi)$ for the interval $[t_{j'}, \bar t_{j'+1})$, both with the same initial condition.
To do that we compare both of them to the auxiliary profile $\phi_2$ generated by the force $b_1$.
From \eqref{phiandphi2}, we have that
\begin{equation}\label{mandphi2}
\|m(\cdot, \bar t_{j'+1})-\phi_2(\cdot,\bar t_{j'+1})\|_{L^2}^2\leq e^{(2+\beta)S}
\frac{\delta_{j'}}{\Delta(\epsilon)}.
\end{equation}
Similarly to \eqref{Itwo} we have:
\[
\frac{d}{dt}\int_{\mathbb{R}} (\phi_2-m^0)^2(x,t)\,dx  =  
\]
\begin{eqnarray*}
& = & 
-2\int_{\mathbb{R}}(\phi_2-m^0)^2(x,t)\,dx+2\int_{\mathbb{R}} b_1(x,t)(\phi_2-m^0)(x,t)\,dx\\
&& +2\int_{\mathbb{R}} (\phi_2-m^0)(x,t)(\tanh(\beta J\ast \phi_2(x,t))-\tanh(\beta J\ast m^0(x,t)))dx\\
&\leq&
C\int_{\mathbb{R}}(\phi_2-m^0)^2(x,t)\,dx+c\int_{\mathbb{R}} \alpha^2 b_1^2(x,t)\,dx,
\end{eqnarray*}
for $c$ large enough.
After applying Gronwall's inequality and \eqref{errc0} we obtain:
\begin{equation}\label{phi2m0}
\|\phi_2(\cdot,\bar t_{j'+1})-m^0(\cdot,\bar t_{j'+1})\|_{L^2}^2
\leq
ce^{(2+\beta)S}\frac{1}{1-c_{*}^{2}C\Delta(\epsilon)}\,\delta_{j'},
\end{equation}
where $\delta_{j'}$ has been defined in \eqref{deltaj}.
Combining \eqref{mandphi2} and \eqref{phi2m0}, for $m$ constructed in \eqref{mwithphi} we have:
\begin{equation}\label {generalbadestimation}
\|m(\cdot,\bar t_{j'+1})-m^0(\cdot,\bar t_{j'+1})\|_{L^2{(\mathbb R)}}^2\leq ce^{(2+\beta)S}\frac{\delta_{j'}}{\Delta(\epsilon)}.
\end{equation}
Moreover, since by the definition of the time $\bar t_{j'+1}$ the profile $m$ is close to $\mathcal M$ at that time, 
we have that
\begin{equation}
	\label{mainestim}
\|\bar m_{\bar\xi(m(\cdot,\bar t_{j'+1}))}-\bar m_{\bar\xi(m^0(\cdot,\bar t_{j'+1}))}\|_{L^2(\mathbb R)}^2
\leq
 ce^{(2+\beta)S}\frac{\delta_{j'}}{\Delta(\epsilon)}+7\vartheta,
\end{equation}
for some $c>0$.
From this, we can obtain an estimate for the distance between the centers in $\bar\xi(m(\cdot,\bar t_{j'+1}))$ 
and $\bar\xi(m^0(\cdot,\bar t_{j'+1}))$.
Let $k$ be the number of centers of $m(\cdot,t_{j'})$ and $\bar r(t_{j'})=(r_1(t_{j'}),...,r_k(t_{j'}))$ with $|r_{i+1}(t_{j'})-r_i(t_{j'})|\geq |\log\epsilon|^2$, $\forall i$.
For $l\in\{1,\ldots, k\}$ odd, define $i_l$ to be the odd label such that
\begin{equation}\label{labels}
\min_{i\,\text{odd}}|\xi_i-\xi^0_l|=|\xi_{i_l}-\xi^0_l|.
\end{equation}
For $l$ even we define $i_l$ analogously.
Furthermore, during the time interval $[t_{j'}, \bar t_{j'+1})$,
new centers might be created due to nucleations. 
Let $\ell_1,\ldots,\ell_p$ be the labels of the newly created centers.

By the properties of the instanton we have that the upper bound in
\eqref{mainestim} induces an upper bound on the volume of the mismatch
between $\bar m_{\bar\xi(\phi(\cdot,\bar t_{j'+1}))}$
and $\bar m_{\bar\xi(m^0(\cdot,\bar t_{j'+1}))}$.
Since the centers $i_1,\ldots, i_k$ are still far enough,
this further induces a bound on the corresponding centers.
Hence, both $|\xi_{i_l}-r_l|$
and $|\xi_{\ell_i}-\xi_{\ell_{i+1}}|$, for $i$ odd in $\{1,\ldots,k\}$ 
are bounded by the estimate in \eqref{mainestim}.

In the next iteration, we construct a profile solving \eqref{mwithphi} for $t\geq \bar t_{j'+1}$ 
starting at $m(\bar t_{j'+1},\cdot)$. Using the same argument as before, 
we choose another time $\bar t_{j'+2} \in[j'+2-\frac 12,j'+2-\frac 14]S$ with
$m(\bar t_{j'+2},\cdot)$ close to $\mathcal M$.
By repeating the same procedure we obtain
\begin{equation}\label {generalbadestimation2}
\|m(\cdot,\bar t_{j'+2})-m^0(\cdot,\bar t_{j'+2})\|_{L^2{(\mathbb R)}}^2\leq ce^{(2+\beta)S}\frac{\delta_{j'+1}}{\Delta(\epsilon)},
\end{equation}
where $m^{0}$ is the solution of the equation without the forcing term in the interval
$[\bar t_{j'+1}, \bar t_{j'+2})$ starting at $m(\cdot,\bar t_{j'+1})$.
This induces a bound on the corresponding centers by the same amount.
These could be the original ones, or the ones nucleated in the time interval $[t_{j'}, \bar t_{j'+1})$ and continued
moving the current one, or those nucleated during the second time interval $[\bar t_{j'+1},\bar t_{j'+2})$.
Thus, during the first two bad time intervals of the connected component $[t_{j'},t_{j''})$, 
the displacement of the old centers
(at time $t_{j'}$) or the distance between the newly created are both bounded by
\[
ce^{(2+\beta)S}\frac{\delta_{j'}}{\Delta(\epsilon)}+ 7\vartheta+
ce^{(2+\beta)S}\frac{\delta_{j'+1}}{\Delta(\epsilon)}+7\vartheta.
\]
At the end of the connected component of the bad time intervals the corresponding estimate is
\begin{equation}\label{bti}
ce^{(2+\beta)S}\frac{1}{\Delta(\epsilon)}\sum_{k=j'}^{j''}\delta_{k}+\frac{P}{\delta(\epsilon)}7\vartheta
\leq 
ce^{(2+\beta)S}\frac{P}{\Delta(\epsilon)}+\frac{P}{\delta(\epsilon)}7\vartheta<<|\log\epsilon|^{2},
\end{equation}
by the choice in \eqref{Delta}.

\section{The particle model, total cost and total displacement}\label{particle}

\subsection{The ``particle" model}

Given a profile $\phi\in\mathcal U[\epsilon^{-1}R,\epsilon^{-2}T]$, in the previous sections
we created a function $m$ with $I(\phi)\geq I(m)$.
By construction, see \eqref{constrofm}, at the end of each good time interval the
function $m$ has its odd/even centers on the
right/left of the corresponding centers of $\phi$, eventually after performing a
jump by a quantity $S_{\epsilon}$ (see \eqref{jumpS2}), if necessary.
To each such center we assign a ``particle" whose position is given
by the function $t\mapsto r_i(t)$ as defined in \eqref{12.200}.
From  \eqref{nstar} there is a maximum possible number of such particles, say $n^*$
and we write $\bar r(t):=(r_1(t),\ldots,r_{n^*}(t))$ for their positions.
During a connected component of good time intervals we may have that some of
these particles die as a result of a ``collision" as described before.
On the other hand, during the bad time intervals (where the cost is higher) we may
get a birth (or more) of two such particles after the occurrence of a nucleation.
Thus, a possible behavior of these particles is the following:
at time $t=0$ we have the particle $r_1(0)=0$ and $r_i(0)=\emptyset$ for all $2\leq i\leq n$, 
which moves in a bad time interval, during which
a nucleation takes place at time $t^*_1\geq 0$ and we have the creation of the new particles 
at positions $r_{i_1}(t^*_1)=r_{i_1+1}(t^*_1)$ (distance $|\log\epsilon|^2$), with $i_1$ odd (note also that we let
$r_{i_1}(t)=r_{i_1+1}(t)=\emptyset$ for $t<t^*_1$). 
Then the particles enter into a connected component of good time intervals after (possibly)
making a jump in their positions $r_i$ by at most $o(|\log\epsilon|^{2})$
as shown in Section~\ref{badtimes}.
Then, before entering into the next good time intervals of small cost,
new jumps may occur as a result of the initialization described in Section~\ref{goodandbad}.
After entering, new jumps have to be taken into account as
a result of a jump from a good time interval to the next as in
Section~\ref{fromgoodtogood}.
In both of these cases (say at a time $t_2^*$) it may happen that two
particles ($r_{i_2}$ and $r_{i_2+1}$) collapse in which case we write
$r_{i_2}(t)=r_{i_2+1}(t)=\emptyset$ for all $t\geq t^*_2$.
Hence, following the above rules and the analysis in the previous sections we obtain the configuration
of the particles denoted by $\{n,(r_1(t),\ldots,r_n(t))\}$ for $t\in [0,\epsilon^{-2}T]$.

\subsection{Lower bound}

We want to find a lower bound of the total cost
determined by the new quantities $\bar{r}(t)$
and the velocities $v_i^0(t)$.
Furthermore, we have the constraint that the total
displacement is $\geq\epsilon^{-1}R$.
From this, we derive a constraint on $v_i^0(t)$, for $t\in [0,\epsilon^{-2}T]$.
We have to take into account the displacement during the good time intervals,
the jumps $S_\epsilon^j$, \eqref{jumpS2}, between two good time intervals, 
the displacement during bad
time intervals \eqref{bti} and finally the displacement due to nucleation and
collision of particles.
Thus, the constraint reads:
\begin{eqnarray}\label{newconstraint}
\sum_{i=1}^{n^*}\int_{\{t:\,r_i(t)\neq\emptyset\}}|v_i^0(t)| & \geq &
\epsilon^{-1}R
-\left(
c n^*\sum_{j\in G_{\rm tot}}\int_{t_j}^{t_{j+1}}(\|\alpha b_1\|_{L^2(d\nu)}^2
+R_{\max})ds\right.
\nonumber\\
&& \left. +c\sum_{j\in G_{\rm tot}}S^j_{\epsilon}
+|\log\epsilon|^2
+n^* 4|\log\epsilon|^2
\right).
\end{eqnarray}

In the good time interval $[t_j,t_{j+1}]$, using \eqref{errc},
we have the following lower bound for the cost:
\[
\int_{t_j}^{t_{j+1}}\int_{\mathbb{R}}\mathcal{H}(\phi,\dot{\phi})(x,t)\,dx\,dt
\geq\int_{t_j}^{t_{j+1}}\|\alpha b_1\|_{L^2(d\nu)}\,dt
-\frac{c_{*}^{2}C\Delta(\epsilon)}{1-c_{*}^{2}C\Delta(\epsilon)}P,
\]
where by H\"older's inequality we also have that
\[
\|\alpha b_1\|_{L^2(d\nu)}
\geq\sum_{i:\,r_i(t)\neq\emptyset}\left\{
\frac{1}{\|\bar m'\|_{L^2(d\nu)}^2}\;
\Big|(\alpha b_1,\bar m'_{\xi_i(t)})_{L^2(d\nu)}\Big|
-c e^{-\alpha|\log\epsilon|^2/2}
\right\}.
\]
Thus, taking also into account the mobility
$\mu=4\|\bar{m}'\|_{L^2(d\nu)}$, in a good time interval we obtain:
\[
\int_{t_j}^{t_{j+1}}\int_{\mathbb{R}}\mathcal{H}(\phi,\dot{\phi})(x,t)\,dx\,dt
\geq
\int_{t_j}^{t_{j+1}}\sum_{i: r_i(t)\ne \emptyset} \frac{ v^0_i(t)^2}{\mu} - c
e^{-\alpha |\log\epsilon|^2/2}2S-
\frac{c_{*}^{2}C\Delta(\epsilon)}{1-c_{*}^{2}C\Delta(\epsilon)}P.
\]
On the other hand,
the cost in a connected component of bad time intervals is neglected unless
if a nucleation occurs. 
Following the notation we used in Section \ref{badtimes}, $[t_{j'}, t_{j''})$ is a generic  connected component of bad time intervals. By using the reversibility property \eqref{revcom} we have that:
\[
\int_{t_{j'}}^{t_{j''}}\int_{\mathbb{R}}\mathcal{H}(\phi,\dot{\phi})(x,t) \,dx\,dt\geq
\mathcal{F}(\phi(\cdot,t_{j''}))-\mathcal{F}(\phi(\cdot,t_{j'})).
\]
Using \eqref{m1.1.000} we have that
for the given $\gamma>0$,
\[
\mathcal{F}(\phi(\cdot,t_{j''}))-\mathcal{F}(\phi(\cdot,t_{j'}))\geq 2q\mathcal{F}(\bar{m})-n^*\gamma,
\]
 where $q$ is the number of nucleations that happened during $[t_{j'},t_{j''}]$.
Thus, for all $\epsilon>0$, the total cost is bounded from below by
\begin{eqnarray}\label{lowerboundcost}
\int_0^{\epsilon^{-2}T}\int_{\mathbb{R}}\mathcal{H}(\phi,\dot{\phi})(x,t)\,dx\,dt & \geq &
\int_{G_{\mathrm{tot}}}\sum_{i: r_i(t)\ne \emptyset} \frac{ v^0_i(t)^2}{\mu} +n\mathcal{F}(\bar{m})-\frac{c_{*}^{2}C\Delta(\epsilon)}{1-c_{*}^{2}C\Delta(\epsilon)}P\nonumber\\
&&
-ce^{-\alpha |\log\epsilon|^2/2}\epsilon^{-2}T
-\gamma,
\end{eqnarray}
 where $n/2$ is the total number of nucleations with $q,\,n\leq n^*$ where $n^*$ is the maximum number of fronts created by the nucleations (see \eqref{nstar}). Thus, the problem reduces to finding the infimum over the velocities $v^0_i(\cdot)$ of the
right hand side of \eqref{lowerboundcost} under the constraint \eqref{newconstraint}, where
$i=1,\ldots,n^*$ is the index of a front and suppose that its lifetime is $T_i$.
With this estimate, arguing as in \cite{ddp} we conclude the proof of the lower bound.

\subsection{Upper bound}

First, we compute the optimal number of nucleations. Then, we construct a sequence
$\phi_{\epsilon}\in \mathcal{U}[\epsilon^{-1}R,\epsilon^{-2}T]$, which at time $t=0$
consists of a multi-instanton with $2n+1$ centers at positions 
$0$ and $\frac{2i}{2n+1}\epsilon^{-1}R\pm
\frac 12|\log\epsilon|^2$, for $i=1,\ldots, n$.
Then for $t\in (0,\epsilon^{-2}T]$ they move with constant velocity $\frac{V}{2n+1}$ to the right (the odd-numbered) or left (the even-numbered), where $V=R/T$. 
When they are at a distance smaller than $|\log\epsilon|^2$ they disappear.
It is easy to check that this sequence satisfies \eqref{upperbound}.

\bigskip

{\bf Acknowledgments.}
We would like to thank Guido Manzi, Nicolas Dirr and Errico Presutti for many fruitful discussions.

\appendix

\section{Existence of solutions of the system \eqref{eqnphi1}-\eqref{eqnm}}\label{ex_uniq}

Recalling the definition of $b$ in \eqref{functionb}
and of $b_{1}$ in \eqref{bone},
we define the sequence $\{\tilde\xi^k, \phi_1^k, m^k\}_{k\geq 1}$ which solves the
following system of equations
(for simplicity we work in the good time interval $[0,S]$):
\begin{eqnarray}
&& b(\phi_1^k)=\alpha_k
b_1,\quad\text{with}\quad
\phi_1^k(\cdot,0)=\phi(\cdot,0)\quad\text{and}\label{sys1}\\
&&b(m^k)=b(\phi_1^k),\quad\text{with}\quad
m^k(\cdot,0)=m_0(\cdot),\label{sys2}
\end{eqnarray} 
where
\begin{equation*}
\alpha_0=1, \quad
\alpha_1=\left(
\frac{1-\bar{m}^2_{\tilde\xi^{0}}}{8}\right)^{\frac{1}{2}}
\quad \text{and}\quad
\alpha_k=\left(
\frac{1-\bar{m}^2_{\tilde\xi^{k-1}}}{8}\right)^{\frac{1}{2}}.
\end{equation*}
The initial condition $m_0$ is as in the initialization in Section~\ref{goodandbad}
and $\tilde\xi^k=(\tilde\xi_1^k,\ldots,\tilde\xi_n^k)$ are the approximate centers of $m^k$ defined as in \eqref{center}.
We define the initial center
$\tilde\xi^0$ as the center of the profile $m^0$, defined by:
\begin{equation*}
b(m^0)=b_1, \quad\text{with\quad}
m^0(\cdot,0)=m_0(\cdot).
\end{equation*}
Then, $m^1$ solves the following initial value problem:
\begin{equation*}
b(m^1)=\alpha_1 b_1, \quad\text{with\quad}
m^1(\cdot,0)=m_0(\cdot).
\end{equation*}
From the equations above for $m^0$ and $m^1$ we have:
\[
\frac{d}{dt}\|m^1(\cdot,t)-m^0(\cdot,t)\|_{L^2}^2
\leq
(2+\beta)\|m^1(\cdot,t)-m^0(\cdot,t)\|_{L^2}^2
+\|(1-\alpha_1)b_1\|_{L^2}^2
\]
But, by the definition of $c_*$ in \eqref{cstar}, it holds that $|(1-\alpha_k)b_1|\leq c_* \alpha_k |b_1|$, for every $k\geq 1$.
Then, applying Gronwall's inequality and using \eqref{errc0}
we obtain:
\begin{equation}\label{app1}
\|m^1(\cdot,t)-m^0(\cdot,t)\|^2_{L^2} \leq 
ce^{(2+\beta) S}\frac{1}{1-c_{*}^2C\Delta(\epsilon)}\int_\mathbb R \int_0^t \mathcal H(x,s)ds\,dx
\leq ce^{(2+\beta) S}\frac{1}{1-c_{*}^2C\Delta(\epsilon)}\delta(\epsilon),
\end{equation}
for some new constant $c>0$.
We define 
\begin{equation}\label{norm}
\|\tilde{\xi}^k-\tilde\xi^{k-1}\|:=\max_{i=1,\ldots,n}|\tilde\xi^k_i-\tilde\xi^{k-1}_i|
\end{equation}
 and estimate
$|\tilde\xi^1_i-\tilde\xi^0_i|$, for $i\in\{1,\ldots,n\}$ by
\[
|\tilde\xi_i^1-\tilde\xi_i^0|\leq c\|m^1-m^0\|_{L^2}.
\]
We first show that $\{\tilde\xi^k\}_{k\geq 0}\subset L^{\infty}([0,S];\mathbb{R}^n)$ is a 
Cauchy sequence. By following the same reasoning as in \eqref{app1}, 
for every $k\geq 1$ 
we have that
\begin{eqnarray}\label{app2}
\|m^k(\cdot,t)-m^{k-1}(\cdot,t)\|_{L^2}^2
& \leq &
ce^{(2+\beta) S}\int_{0}^{t}\|b(m^k)(\cdot,s)-b(m^{k-1})(\cdot, s)\|_{L^2}^2ds\nonumber\\
& \leq &
ce^{(2+\beta) S}2\frac{1}{1-c_{*}^2C\Delta(\epsilon)}\delta(\epsilon).
\end{eqnarray}
Therefore, since $\|m^k-m^{k-1}\|_{L^2}$ is small, 
given a mixed contour $\Gamma_i$ we have that: 
\begin{equation}\label{app3}
|\tilde\xi^k_i-\tilde\xi^{k-1}_i|\leq C\|m^k-m^{k-1}\|_{L^2}.
\end{equation}
For the difference between the two forces $b(m^k)$ and $b(m^{k-1})$, from \eqref{sys1}
and \eqref{sys2} we have:
\begin{eqnarray}\label{app4}
\int_{0}^{t}\|b(m^k)-b(m^{k-1})\|_{L^2}^2ds&=&\int_{0}^{t}\int_{\mathbb{R}}\left(\left(
\frac{1-\bar{m}^2_{\tilde\xi^{k-1}}}{8}\right)^{\frac{1}{2}}
-\left(
\frac{1-\bar{m}^2_{\tilde\xi^{k-2}}}{8}\right)^{\frac{1}{2}}
\right)^2b_1(x,s)^2dx\,ds \nonumber \\
&\leq&\frac 1 8\int_{0}^{t}\int_{\mathbb{R}}|\bar{m}^2_{\tilde\xi^{k-1}}
-\bar{m}^2_{\tilde\xi^{k-2}}|\,b_1(x,s)^2dx\,ds \nonumber\\
&\leq&\frac{ (\Delta(\epsilon))^2 }{4}\sum_{i=1}^{n}\int_{0}^{t}\int_{\Gamma_{i}}|\bar{m}_{\tilde\xi^{k-1}}
-\bar{m}_{\tilde\xi^{k-2}}|\mathbf{1}_{[|b(\phi)|\leq \Delta(\epsilon)]}dx\,ds \nonumber\\
&\leq&\frac{ (\Delta(\epsilon))^2 }{2}nS\|\bar m '\|_{L^1}\sup_{0\leq s\leq t}\|\tilde\xi^{k-1}-\tilde\xi^{k-2}\|(s)
\end{eqnarray}
In the above computations we 
exploited the fact that $m^k$ and $m^{k-1}$ have the same number of contours and their centers are close to each other due to \eqref{app3}. 
We combine \eqref{app2}, \eqref{app3}, \eqref{app4} and for 
$\epsilon$ sufficiently small we obtain a contraction:
\[
\sup_t\|\tilde\xi^{k}-\tilde\xi^{k-1}\|\leq L\sup_t \|\tilde\xi^{k-1}-\tilde\xi^{k-2}\|
\]
where $L=C\|\bar{m}'\|_{L^1}e^{\beta S}\Delta^2nS<1$.

Similarly, using the same estimates
we can show that the sequences $\{m^k\}_k$ and $\{\phi_1^k\}_k$ are Cauchy
in the norm $\sup_t(\|\cdot\|_{W^{1,1}})$ and using a standard argument
we can show that the limit point satisfies the system.

\section{$L^1$ and $L^{2}$ bounds on the centers}\label{L1bound}

We denote
\[
\mathcal{N}=\{
m\in L^{\infty}(\mathbb{R},[-1,1]): \limsup_{x\to -\infty}m(x)<0;\,\,
\liminf_{x\to +\infty}m(x)>0
\}
\]
and define the $\delta$ neighborhood of 
$\mathcal{M}^{(1)}:=\{\bar{m}_{\xi},\,\xi\in\mathbb{R}\}$ by
\[
\mathcal{M}^{(1)}_{\delta}=\bigsqcup_{\xi\in\mathcal{R}}\{m\in L^{\infty}
(\mathcal{R},[-1,1]):\,\|m-\bar{m}_{\xi}\|_{L^2}<\delta
\}.
\]

\begin{lemma}\label{L1center}
Any $m\in\mathcal{N}$ has a center.
Moreover, there are positive constants $c$ and $\delta$ so that any 
$m\in\mathcal{M}^{(1)}_\delta$ has a unique center $\xi(m)$.
Furthermore, for any $n\in\mathcal{M}^{(1)}_{\delta}$ with $\|m-n\|_{L^1}$ small
we have:
\[
|\xi(m)-\xi(n)|\leq c\|m-n\|_{L^1}.
\]
The same result also holds for the $\|\cdot\|_{L^{2}}$ norm.
\end{lemma}

{\it Proof:}
From the definition of a center it suffices to find a $\xi$  such
that 
\begin{equation}\label{center_append}
(m,\bar{m}_{\xi}')_{L^2(d\nu_{\xi})}=0
\end{equation}
The function $\xi\mapsto (m,\bar{m}_{\xi}')_{L^2(d\nu_{\xi})}$ is a continuous
function and by the definition of $\mathcal{N}$ we have that
\[
\limsup_{x\to -\infty}(m,\bar{m}_{\xi}')_{L^2(d\nu_{\xi})}<0;\,\,
\liminf_{x\to +\infty}(m,\bar{m}_{\xi}')_{L^2(d\nu_{\xi})}>0.
\]
Thus \eqref{center_append} has a solution.

To show uniqueness, since the function $m$ is in the $\delta$-ball 
around some $\bar{m}_{\xi_0}$
(without loss of generality we can also assume that $\xi_0=0$), we write
\[
m=\bar{m}+\psi,\,\,\|\psi\|_{L^2(d\nu)}<\delta.
\]
Then \eqref{center_append} gives 
$(m,\bar{m}_{\xi}')_{L^2(d\nu_{\xi})}=-(\psi,\bar{m}_{\xi}')_{L^2(d\nu_{\xi})}$
and since $\|\psi\|_{L^2(d\nu)}<\delta$,
we obtain that
\begin{equation}\label{center2}
|(\psi,\bar{m}_{\xi}')_{L^2(d\nu_{\xi})}|\leq
\|\psi\|_{L^2(d\nu)}\|\bar{m}_{\xi}'\frac{1-\bar{m}^2}{1-\bar{m}_{\xi}^2}\|_{L^2(d\nu)}\leq
\delta\frac{1}{1-m_{\beta}^2}\|\bar{m}'\|_{L^2(d\nu)},\,\,\,\mathrm{for\, any}\, \,\xi\in\mathbb{R}.
\end{equation}
Following \cite{errico}, Theorem 8.5.1.1,
we choose $\delta<\frac{\alpha_0}{\|\bar{m}_{\xi}'\|_{L^2(d\nu_{\xi})}}$
which implies that there is no solution to \eqref{center2} when $|\xi|\geq 1$
and $\|m-\bar{m}\|_{L^2(d\nu)}<\delta$.

Given $n$ with $\|m-n\|_{L^1}$ small, we write: $
n=m+\chi, \,\,\mathrm{with}\,\,\|\chi\|_{L^1}<\delta'
$.
We define
\begin{equation}\label{gxi}
g(\xi):=(\bar{m},\bar{m}_{\xi}')_{L^2(d\nu_{\xi})}
+(\psi,\bar{m}_{\xi}')_{L^2(d\nu_{\xi})}
+(\chi,\bar{m}_{\xi}')_{L^2(d\nu_{\xi})}
\end{equation}
Then $\xi(n)$ is defined by $g(\xi(n))=0.$ We have:
\[
0=g(\xi(n))=(\chi,\bar{m}_{\xi(m)}')_{L^2(d\nu_{\xi(m)})}+
\int_{\xi(m)}^{\xi(n)}g'(z)dz
\]
Since $|\xi(n)|\leq 1$ and $|\xi(m)|\leq 1$ we have that $|z|\leq 1$,
thus $g'(z)\geq\alpha_0/2$.
Hence,
\[
|\xi(n)-\xi(m)|
\leq\frac{2}{\alpha_0}|(\chi,\bar{m}_{\xi(m)}')_{L^2(d\nu_{\xi(m)})}|
\leq\frac{2}{\alpha_0}\|\chi\|_{L^1}\|\frac{\bar{m}_{\xi(m)}'}{1-\bar{m}^2_{\xi(m)}}\|_{\infty}
\]
which concludes the proof. 
Alternatively, we can have the following inequality:
\[
|\xi(n)-\xi(m)|
\leq\frac{2}{\alpha_0}|(\chi,\bar{m}_{\xi(m)}')_{L^2(d\nu_{\xi(m)})}|
\leq\frac{2}{\alpha_0}\|\chi\|_{L^2(d\nu)}\|\bar{m}_{\xi(m)}'\|_{L^{2}(dx)},
\]
which concludes the proof for the case of the $L^{2}$ norm as well.
\qed

\section{Asymptotic analysis of $\mathcal{H}$}\label{bsquare}

For $\mathcal H$ given in \eqref{mathcalH} we have that
uniformly on $u\in[-1,1]$ and $w\in(-1,1)$:
\begin{equation*}
\lim_{|b|\to\infty}\frac{\mathcal{H}(b,u,w)}{|b|\log(|b|+1)}=
\frac{1}{2}\qquad\mathrm{and}
\qquad
\lim_{|b|\to 0}\frac{\mathcal{H}(b,u,w)}{b^2}=
\frac{1}{4(1+uw)}.
\end{equation*}
Moreover, for the choice of $\Delta(\epsilon)$ in \eqref{Delta}, in the case $|b|\leq\Delta(\epsilon)$,
we have that:
\[
|\mathcal{H}(b,u,w) -\frac{1}{4(1+u w))}b^2|\leq C\,|b|^3\leq  C\,\Delta(\epsilon)^3,
\]
for some $C>0$.
Thus, for $b_{1}$ defined in \eqref{bone}, using \eqref{cstar} we have that for the same constant $C>0$
the following hold:
\begin{equation}\label{errc0}
\int_{\{|b|\leq\Delta(\epsilon)\}}|\alpha(x,t) b_1(x,t)|^2 dx\,dt
\leq 
\frac{1}{1-c_{*}^{2}C\Delta(\epsilon)}\int_{\{|b|\leq\Delta(\epsilon)\}}\mathcal{H}(b,u,w)dx\,dt
\end{equation}
and
\begin{eqnarray*}
\int_{\{|b|\leq\Delta(\epsilon)\}} \big|\mathcal{H}(b,u,w) -\frac{1}{4(1+u w)}b_1^2\big|dx\,dt
&\leq & C\Delta(\epsilon)\int_{\{|b|\leq\Delta(\epsilon)\}}b^{2}(x,t)\,dx\,dt\nonumber\\
&\leq & c_{*}^{2}C\Delta(\epsilon)\int_{\{|b|\leq\Delta(\epsilon)\}}|\alpha(x,t) b(x,t)|^{2}\,dx\,dt.
\end{eqnarray*}
Adding and subtracting $\int_{\{|b|\leq\Delta(\epsilon)\}} \mathcal{H}(b,u,w)dx\,dt$,
for $\epsilon$ small enough
it is further implied that
\begin{equation}\label{errc}
\int_{\{|b|\leq\Delta(\epsilon)\}} \big|\mathcal{H}(b,u,w) -\frac{1}{4(1+u w)}b_1^2\big|dx\,dt
\leq  \frac{c_{*}^{2}C\Delta(\epsilon)}{1-c_{*}^{2}C\Delta(\epsilon)} 
\int_{\{|b|\leq\Delta(\epsilon)\}} \mathcal{H}(b,u,w)dx\,dt,
\end{equation}
which is small as $\epsilon\to 0$ since the cost is bounded by $P$
and $\Delta(\epsilon)\to 0$.

\bigskip
\bigskip

\end{document}